\documentclass[%
 reprint,
superscriptaddress,
 amsmath,amssymb,
 aps,
prm,
]{revtex4-1}

\usepackage{graphicx}% Include figure files
\usepackage{epstopdf}
\usepackage{dcolumn}% Align table columns on decimal point
\usepackage{bm}% bold math
\usepackage{gensymb}
\usepackage{upgreek}
\usepackage{hyperref}
\usepackage{tabularx}

\preprint{APS/123-QED}

\begin{document}

\title{Hydroflux-Controlled Growth of Magnetic K-Cu-Te-O(H) Phases}
\author{Allana G. Iwanicki}
\thanks{These authors contributed equally}
\affiliation{Department of Chemistry, Johns Hopkins University, 3400 N. Charles Street, Baltimore, MD 21218, United States of America}
\affiliation{Institute for Quantum Matter, William H. Miller III Department of Physics and Astronomy, Johns Hopkins University, 3400 N. Charles Street, Baltimore, MD 21218, United States of America}
\author{Brandon Wilfong}
\thanks{These authors contributed equally}
\affiliation{Department of Chemistry, Johns Hopkins University, 3400 N. Charles Street, Baltimore, MD 21218, United States of America}
\affiliation{Institute for Quantum Matter, William H. Miller III Department of Physics and Astronomy, Johns Hopkins University, 3400 N. Charles Street, Baltimore, MD 21218, United States of America}
\author{Eli Zoghlin}
\affiliation{Institute for Quantum Matter, William H. Miller III Department of Physics and Astronomy, Johns Hopkins University, 3400 N. Charles Street, Baltimore, MD 21218, United States of America}
\author{Wyatt Bunstine}
\affiliation{Institute for Quantum Matter, William H. Miller III Department of Physics and Astronomy, Johns Hopkins University, 3400 N. Charles Street, Baltimore, MD 21218, United States of America}
\author{Maxime A. Siegler}
\affiliation{Department of Chemistry, Johns Hopkins University, 3400 N. Charles Street, Baltimore, MD 21218, United States of America}
\author{Tyrel M. McQueen}
\affiliation{Department of Chemistry, Johns Hopkins University, 3400 N. Charles Street, Baltimore, MD 21218, United States of America}
\affiliation{Institute for Quantum Matter, William H. Miller III Department of Physics and Astronomy, Johns Hopkins University, 3400 N. Charles Street, Baltimore, MD 21218, United States of America}
\affiliation{Department of Materials Science and Engineering, Johns Hopkins University, 3400 N. Charles Street, Baltimore, MD 21218, United States of America}

\date{\today}

\begin{abstract}
Innovative synthetic approaches can yield new phases containing novel structural and magnetic motifs. In this work, we show the synthesis and magnetic characterization of three new and one previously reported layered phase in the K-Cu-Te-O(H) phase space using a tunable hydroflux technique. The hydroflux, with a roughly equal molar ratio of water and alkali hydroxide, is a highly oxidizing, low melting solvent which can be used to isolate metastable phases unattainable through traditional solid state or flux techniques. The newly synthesized phases, K$_{2}$Cu$_{2}$TeO$_{6}$, \mbox{K$_{2}$Cu$_{2}$TeO$_{6}$ $\cdot$ H$_{2}$O}, and \mbox{K$_{6}$Cu$_{9}$Te$_{4}$O$_{24}$ $\cdot$ 2 H$_{2}$O}, contain Cu$^{2+}$ within CuO$_{4}$ square planar plaquettes and TeO$_{6}$ octahedra ordering to form structural honeycomb layers isolated by interlayer K$^{+}$ ions and H$_{2}$O molecules. We find the synthesized structures display varying tilt sequences of the CuO$_{4}$ plaquettes, leading to distinct Cu$^{2+}$ magnetic motifs on the structural honeycomb lattice and a range of effective magnetic dimensionalities. We find that \mbox{K$_{2}$Cu$_{2}$TeO$_{6}$ $\cdot$ H$_{2}$O} does not order and displays alternating chain Heisenberg antiferromagnetic (AFM) behavior, while K$_{2}$Cu$_{2}$TeO$_{6}$ and \mbox{K$_{6}$Cu$_{9}$Te$_{4}$O$_{24}$ $\cdot$ 2 H$_{2}$O} order antiferromagnetically (T$_{N}$ = 100 K and T$_{N}$ = 6.5 K respectively). The previously known phase, \mbox{K$_{2}$CuTeO$_{4}$(OH)$_{2}$ $\cdot$ H$_{2}$O}, we find contains structurally and magnetically one-dimensional CuO$_{4}$ plaquettes leading to uniform chain Heisenberg AFM behavior and shows no magnetic order down to T = 0.4 K. We discuss and highlight the usefulness of the hydroflux technique in novel syntheses and the interesting magnetic motifs that arise in these particular phases.

\end{abstract}

\maketitle

\section{Introduction}

\par Structure-property relationships are the foundation of solid state chemistry and physics. However, synthesis-structure relationships often remain mysterious due to new combinations of elements present, extreme environments required, and the scientists’ inability to observe the reactions in real time. In the case of flux-based growths, the structure of the synthesized crystal is tied to the identity and stability of the ionic fragments present in solution.\cite{Rabenau1985, Elwell1976} Thus, to rationally synthesize new materials via flux-based reactions, conditions must be tuned to adjust the stability of different ionic building blocks in solution in order to stabilize the desired phases. 

\par To this end, we explore the use of hydroflux reactions as a method to access underexplored regions of phase space via the unique chemical conditions of these fluxes. In general, hydroflux reactions use concentrated, wet hydroxides as a basic reaction medium heated to moderate temperatures in a sealed autoclave in a cross between hydrothermal and hydroxide flux methods.\cite{Chance2014} However, relative to hydrothermal and flux reactions, the unique chemistry which occurs in hydroflux reactions due to their nearly equal amounts of water molecules and hydroxide ions is underexplored. In particular, ionic effects suppress water vaporization and pressure buildup in the reaction chamber, and highly reactive peroxides, superoxides, and oxygen ions are formed in solution.\cite{Lux1959, Chance2014} These factors create unique growth conditions that can yield novel oxidized phases at low temperatures and pressures. Like hydrothermal and hydroxide flux syntheses, hydroflux syntheses are highly tunable and different products may be obtained by adjusting solution or reagent concentrations, identity of hydroxide, and synthesis time.\cite{Zhou2022, Rabenau1985, Chance2014} Product dimensionality has also been shown to be targetable in mixed hydroxide/halide fluxes by tuning the solubility of the reagents and thus the ratios and identities of constituent building blocks in solution.\cite{Zhou2022} Here, we seek to explore these aspects of hydroflux syntheses by targeting products with desirable building blocks and dimensionality for hosting exotic magnetic phases. 

\begin{figure*}[th]
    \centering
    \includegraphics[height = 3in, width = 5.5in]{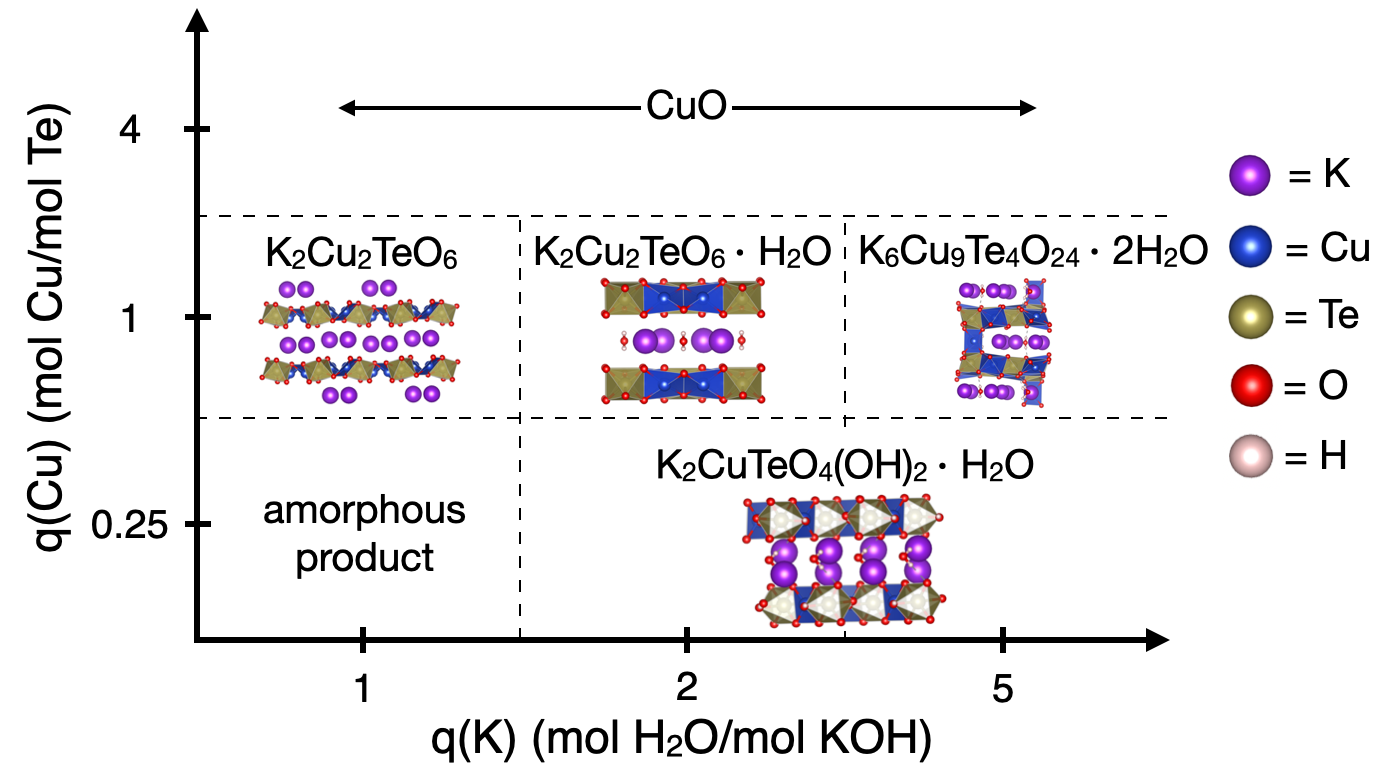}
    \caption{Schematic of all isolated products formed from hydroflux reactions in the K-Cu-Te-O(H) phase space depending on the q(K) = (mol H$_{2}$O/ mol KOH) and q(Cu) = (mol Cu/ mol Te) used. All reactions were washed with water to isolate the products.}
    \label{fig:hydroflux_overview}
\end{figure*}

\par In this work, we used the hydroflux method to target novel phases in the Cu-Te-O(H) phase space containing fully oxidized Cu$^{2+}$, a model spin-$\frac{1}{2}$ ion. Fully oxidized Te$^{6+}$, which strongly prefers octahedral coordination, can form a stable, non-magnetic scaffold around Cu$^{2+}$\cite{Christy2016} to create model magnetic systems such as 1-dimensional (1D) chains, dimers, and honeycomb motifs.\cite{Norman2018, Rosner2007} Additionally, incorporating an alkali hydroxide like KOH could further stabilize new structures as large alkali cations can partition Cu-Te-O(H) layered building blocks and tune the dimensionality of the systems.\cite{Xu2005} Generally, layered phases containing magnetic ions can exhibit exotic physical phenomena due to the competing effects of electronic confinement to specific low dimensional geometries and magnetic interactions within them.\cite{Balents2010} Multiple competing orders on a single lattice can drive slight structural changes, yield unusual orbital overlap, and encourage collective electronic states.\cite{Da2014, Chamorro2020}

\par This work shows the directed synthesis of new phases based on guiding principles for hydroflux growth techniques. We synthesized four new phases in the K-Cu-Te-O(H) phase space and found that KOH hydrofluxes result in K incorporation and the formation of 2D layered Cu-Te-O(H) frameworks, isolated by intercalated K$^{+}$ ions and H$_{2}$O molecules.  All synthesized products are shown in Fig. \ref{fig:hydroflux_overview}. These new phases have interesting Cu$^{2+}$ motifs, including honeycomb-like layers and isolated chains. We have characterized the magnetic properties of these various Cu$^{2+}$ motifs, revealing magnetic order as well as uniform and alternating spin chain behaviors. Overall this work has broad implications for the synthetic design of new phases using hydroflux growth as well as the study of fundamental magnetic properties of spin-$\frac{1}{2}$ systems with different magnetic motifs.

\section{Experimental Methods}

Samples were synthesized via hydroflux reactions as follows: powder reagents CuO (Thermo Scientific, 99.995\%) and TeO$_{2}$ (Acros Organics, 99+\%) were combined with a total quantity of 1-3 mmol  to achieve molar ratios defined as \mbox{q(Cu) = (moles Cu)/(moles Te) = 0.25, 1, 4}. KOH (Fisher Chemical 86.6\%), deionized (DI) H$_{2}$O, and H$_{2}$O$_{2}$ (Fisher Chemical 30\%) were combined to achieve molar ratios defined as \mbox{q(K) = (moles H$_{2}$O)/(moles KOH) = 1, 2, 5}, with fixed liquid reagent amounts of 2 mL DI H$_{2}$O and 1 mL H$_{2}$O$_{2}$. q(K) was calculated assuming anhydrous KOH for simplicity, and 1 g H$_{2}$O$_{2}$ = 0.7 g H$_{2}$O. Reagents were loaded into a 22 mL capacity teflon-lined autoclave with H$_{2}$O$_{2}$ added last and dropwise to minimize sudden O$_{2}$ gas formation. The autoclaves were placed in stainless steel containers, closed, and heated at 200 $^{\circ}$C for between 2 and 4 days in a low temperature oven before being quenched to room temperature. Samples were rinsed with DI H$_{2}$O and filtered using a vacuum funnel. \mbox{K$_{2}$Cu$_{2}$TeO$_{6}$} formed large, flat, emerald green, hexagonal crystals. \mbox{K$_{2}$Cu$_{2}$TeO$_{6}$ $\cdot$ H$_{2}$O} formed point-centered clusters of flat, clover-shaped kelly green crystals. \mbox{K$_{6}$Cu$_{9}$Te$_{4}$O$_{24}$ $\cdot$ 2 H$_{2}$O} formed small forest green crystals interspersed with teal polycrystalline material (both grown on top of large black crystals). \mbox{K$_{2}$CuTeO$_{4}$(OH)$_{2}$ $\cdot$ H$_{2}$O} formed dark green-black crystals with several growth faces. Images of the crystals are shown in the SI. Single crystal X-ray diffraction (SCXRD) measurements were performed using a SuperNova diffractometer (equipped with Atlas detector) with Mo K$\alpha$ radiation ($\lambda$ = 0.71073 \AA) under the program CrysAlisPro (Version 1.171.42.49, Rigaku OD, 2020 - 2022). The same program was used to refine the cell dimensions and for data reduction. All reflection intensities were measured at T = 213(2) K. The structure was solved with the program SHELXS-2018/3 and was refined on \textit{F}$^{2}$ with SHELXL-2018/3.\cite{Sheldrick2008} Analytical numeric absorption corrections using a multifaceted crystal model were performed using CrysAlisPro. The temperature of the data collection was controlled using the Cryojet system (Oxford Instruments). Phase purity was determined using powder X-ray diffraction (pXRD) on a Bruker D8 Focus diffractometer equipped with a LynxEye detector using Cu K$\alpha$ radiation ($\lambda$ = 1.5406 \AA). Data was collected in the range 2$\theta$ = 5-120$^{\circ}$ with a step size of 0.01715$^{\circ}$ and a step time of 2 seconds. pXRD Rietveld refinements were performed using Topas5 using the refined single crystal structure as the starting point refinement for each compound.\cite{Perl2012} Subsequently, only lattice parameters, peak shape, and instrumental zero error were refined and changed from the single crystal solution. Temperature-dependent magnetic susceptibility data was collected on a Quantum Design Magnetic Property Measurement System (MPMS3) from T = 2 - 300 K under an applied field of H = 100 Oe and $\mu_{0}H$ = 1 T using zero-field-cooled (ZFC) and field-cooled (FC) protocols. To measure down to T = 0.4 K, the $^{3}$He option of the MPMS system was used and the DC magnetic susceptibility was measured in a field H = 100 Oe in the T = 0.4 - 2 K temperature range. Isothermal magnetization measurements were collected at various temperatures with a field range of $\pm$7 T. All magnetic data were collected on powderized single crystal samples. Scanning electron microscopy (SEM) and energy dispersive spectroscopy (EDS) were performed using a JEOL JSM-IT100 by mounting single crystals on carbon tape.  All crystal structure visualizations were done using VESTA.\cite{Momma2008}

\section{Hydroflux synthesis}

\begin{table*}[th]
    \centering
\caption{Single crystal X-ray diffraction data of all compounds synthesized in this work: K$_{2}$C\lowercase{u}$_{2}$T\lowercase{e}O$_{6}$, K$_{2}$C\lowercase{u}$_{2}$T\lowercase{e}O$_{6}$ $\cdot$ H$_{2}$O, K$_{6}$C\lowercase{u}$_{9}$T\lowercase{e}$_{4}$O$_{24}$ $\cdot$ 2 H$_{2}$O, K$_{2}$C\lowercase{u}T\lowercase{e}O$_{4}$(OH)$_{2}$ $\cdot$ H$_{2}$O} 
\begin{tabular}{lcccc}
\hline\hline
Compound                      & K$_{2}$C\lowercase{u}$_{2}$T\lowercase{e}O$_{6}$ & K$_{2}$C\lowercase{u}$_{2}$T\lowercase{e}O$_{6}$ $\cdot$ H$_{2}$O & K$_{6}$C\lowercase{u}$_{9}$T\lowercase{e}$_{4}$O$_{24}$ $\cdot$ 2 H$_{2}$O & K$_{2}$C\lowercase{u}T\lowercase{e}O$_{4}$(OH)$_{2}$ $\cdot$ H$_{2}$O \\
T (K)                         & 213(2)              & 213(2)              & 213(2)              & 213(2) \\
Space group                   & $P2_{1}/c$ (14)             & $Cmcm$  (63)            & $Pmn2_{1}$  (31)            & $Cc$ (9)       \\
Crystal system                & monoclinic        & orthorhombic        & orthorhombic        & monoclinic\\
$a$ (\AA)                     &  6.4009(2)         & 8.7289(4)          &  	12.6180(4)            & 9.5773(3)       \\
$b$ (\AA)                     & 9.2339(2)          & 5.8069(3)           & 10.5960(3)           &  6.22756(18)    \\
$c$ (\AA)                     & 5.28005(17)           & 12.7410(6)           & 9.2348(2)           & 11.9847(4)      \\
$\beta$ ($^{\circ}$)                     &  104.647(3)           & 90         & 90          &  90.493(3)        \\
Volume (\AA$^{3}$)            &  	301.937(16)            &  	645.81(5)           & 1234.70(6)            &  	714.78(4)  \\
Z                             & 2                   & 2                   & 2                   & 2          \\
$\rho_{calc}$  (g/cm$^{3}$)   & 4.717              &  4.596                & 4.672               & 3.581      \\
$\lambda$, Mo K$\alpha$ (\AA) & 0.71073             & 0.71073             & 0.71073             & 0.71073      \\
No. reflections collected     & 11030               & 9294               &  27504                 & 10524          \\
No. independent reflections   & 1132                 & 662                 &  4784                 & 2686           \\
2$\theta$ range ($^{\circ}$)  &  6.58 to 65.984        & 6.396 to 65.958       & 3.844 to 65.996        &  6.8 to 65.996  \\     
Index ranges                  & $ -9 \leq h \leq 9$ & $ -13 \leq h \leq 13$ & $ -19 \leq h \leq 19$ & $ -14 \leq h \leq 14$ \\
                              & $ -14 \leq k \leq 14$ & $ -8 \leq k \leq 8$ & $ -16 \leq k \leq 16$ & $ -9 \leq k \leq 9$ \\
                              & $ -8 \leq l \leq 8$ & $ -19 \leq l \leq 19$ & $ -14 \leq l \leq 14$ & $ -18 \leq l \leq 18$ \\
Data/Restraints/Parameters    & 1132/0/53            & 662/1/37            & 4784/1/218           &  2686/7/113    \\
$F(000)$                      & 392.0              &  824.0              &  1590               & 716        \\
Goodness-of-fit on F$^{2}$ & 1.193 & 1.194 & 1.058 & 1.153 \\
$R_{1}, wR_{2}$  [I $\geq$ 2$\sigma$(I)] & 0.0108, 0.0241  & 0.0199, 0.0524  & 0.0250,  0.0451     & 0.0143, 0.0375 \\ 
$R_{1}, wR_{2}$  [all data]   & 0.0112, 0.0242       & 0.0219, 0.0533       & 0.0313, 0.0477     & 0.0148, 0.0377\\ 
Largest diff. peak/hole (/e\AA$^{-3}$) & 0.50/-0.58  & 1.47/-0.74           &  	1.47/-0.90         & 0.47/-0.87  \\ \hline\hline
\end{tabular}
\label{scxrd}
\end{table*}

\par Hydroflux synthesis is a relatively new and underexplored type of reaction which sits between the well-studied regimes of hydrothermal and hydroxide flux reactions. Specifically, hydroflux reactions occur in molten, wet hydroxides with a roughly equal molar ratio of hydroxide to water. Chemically, this means that each water molecule is loosely bound to a hydroxide molecule which limits its activity,\cite{Albrecht2021} in contrast to the water and hydroxide environments of hydrothermal and hydroxide flux syntheses. In order to explain the usefulness and novelty of hydroflux reactions, we will briefly compare them to hydrothermal and hydroxide flux reactions. 

\par In hydrothermal reactions, when water is the primary solvent, H$_{2}$O autodissocates to form OH$^{-}$ and H$^{+}$ ions. This dissociation is typically controlled by tuning the pH of the solution. Additionally, based on the strength and polarization of cation-anion bonds in the reagents, the OH$^{-}$ or H$^{+}$ can effectively break the bonds in the reagents and dissolve the species.\cite{Flood1947} Heating water above its boiling point in a closed system, even below its supercritical temperature, enhances the activity of both the OH$^{-}$ and H$^{+}$ ions and enables dissolution of a wide range of reagents into ionic fragments.\cite{wilfong2021} In conjunction, the dielectric constant of water under pressure is decreased with increasing temperature, resulting in some species precipitating out of solution at high temperatures.\cite{Rabenau1985} Additional species may be added to modify the pH, as a mineralizer to encourage precipitation or dissolution, or change redox potentials.\cite{wilfong2021} Depending on these factors, metals may prefer to exist in a range of coordination environments, bound to several ligands, and with the particular coordination geometry determined by the electron count of the ligand(s) and polarization of the bonds. Temperature gradients and/or the activity of mineralizers in solution can cause fragments to pair and precipitate out of solution; the geometries of metal ions that can be stabilized in solution can remain when the fragments form an extended solid.\cite{Elwell1976} One key additional benefit of hydrothermal reactions is their low temperatures relative to traditional solid-state synthesis methods, which can be exploited to limit the effects of thermodynamics and drive the stabilization of metastable phases.\cite{wilfong2021}

\par Hydroxide flux reactions are often performed in open crucibles above the melting point of the dry hydroxide. Hydroxides are highly ionic species which tend to separate into cations and hydroxide anions and are oxidizing at elevated temperatures.\cite{Elwell1976} The strength of the autodissociation depends on the identity of the cation, and the hydroxide anions can further dissociate into H$_{2}$O and O$^{2-}$.\cite{Albrecht2021} Beyond these autodissociation reactions, molten alkali hydroxides have been found to produce oxidizing peroxides and superoxides in the presence of oxygen; heavier cations tend to produce more of these species.\cite{Lux1959} Thus, the ability of hydroxide flux reactions to dissolve and precipitate species from solution depends strongly on the identity of the hydroxl counterion. In contrast to hydrothermal reactions, hydroxide flux reactions are performed at high temperatures and tend to produce the most oxidized phase.\cite{Elwell1976} In general, hydroxide flux reactions can be particularly useful when considering reactions where the solubility of reagents are low in aqueous or molten salt environments and when strongly oxidizing conditions are needed to stabilize desired oxidation states. 

\par The hydroflux technique differs from its peer synthetic methods in several key ways. Unlike hydrothermal and solid-state techniques, the hydroflux scheme is strongly oxidizing and enables the controlled formation of phases with high oxidation states. For example, similarly oxidizing hydroxide fluxes have been used to reduce extreme oxygen pressure requirements in the synthesis of high valent nickelates.\cite{Wang2023, Dicarlo1994} As in hydroxide flux reactions, the identity of the hydroxide in hydroflux affects the basicity of the melt and the equilibria which dictate formation of peroxides and superoxides. In hydroflux reactions, the presence of water likely changes these equilibria, resulting in different products. Water affects the local coordination of metal ions in solution, as the presence of comparable amounts of stronger field water and weaker field hydroxide ligands can influence preferred geometries.\cite{harms1986} In this work, water-rich hydrofluxes allowed hydroxl groups and water to be incorporated into the structures, while strict hydroxide flux (and, in this work, hydroxide-rich hydroflux) syntheses tend to form oxides.\cite{Albrecht2021} Furthermore, while flux reactions, compared to solid-state syntheses, are often championed as low-temperature methods capable of stabilizing metastable products, hydrofluxes can have melt (and thus reaction) temperatures several hundred degrees below their respective hydroxides. Hydroflux reactions may be performed at temperatures closer to those in hydrothermal reactions, allowing the exploration of metastable phases in an oxidizing environment. Finally, the strongly oxidizing hydroflux and hydroxide flux reactions have also been used to solubilize reagents insoluble in both aqueous media and other fluxes.\cite{roof2010} Overall, the use of hydrofluxes as opposed to traditional hydrothermal or flux reactions enables novel regions of phase space to be explored due to the pairing of highly oxidizing and low-temperature conditions with the unique dissolution properties of a mixed water/hydroxide flux. 

\par Here, we give a synopsis of the structures reported in this work in the context of the changes in the hydroflux reaction scheme and the resulting changes in structure to help elucidate the chemical considerations in product selection from hydroflux reactions. The structure and SCXRD data for the four phases stabilized in this work are shown in Table \ref{scxrd} and the SI. 

\par We have found three new phases (K$_{2}$Cu$_{2}$TeO$_{6}$, \mbox{K$_{2}$Cu$_{2}$TeO$_{6}$ $\cdot$ H$_{2}$O}, and \mbox{K$_{6}$Cu$_{9}$Te$_{4}$O$_{24}$ $\cdot$ 2 H$_{2}$O}) in the q(Cu) = 1 regime with q(K) = 1, 2, and 5 respectively. For all q(Cu) = 1 products, the ratio of Cu:Te in the product remains fixed at $\approx$ 2, implying that, within the explored regime of q(K), the ratio of solubility of CuO to TeO$_{2}$ does not change. We do find that as q(K) is increased with q(Cu) = 1, the amount of hydrate in the product increased demonstrating that for higher q(K) more H$_{2}$O remains active in solution. Structurally, the layered arrangements of CuO$_{4}$ square planar plaquettes and TeO$_{6}$ octahedra form a honeycomb motif with different interlayer spacer species. In all of these phases, the Cu-O coordination environment can be effectively treated as square planar as CuO$_{6}$ octahedra show distortion via large apical elongation. Although all of these phases appear structurally similar, slight differences in the directionality of the octahedral distortions result in vastly different magnetic interactions on the structural honeycomb motif. 

\par After the successful synthesis of three new phases in the q(Cu) = 1 regime, the ratio of Cu:Te in the starting material was systematically modified. When q(Cu) is increased to 4 for all q(K) tested, CuO is recovered as the majority phase and no extended Cu-Te-O(H) motifs are formed. Additionally, when q(Cu) is decreased to 0.25, the only phase stabilized is the previously reported \mbox{K$_{2}$CuTeO$_{4}$(OH)$_{2}$ $\cdot$ H$_{2}$O} phase.\cite{Effenberger1993} \mbox{K$_{2}$CuTeO$_{4}$(OH)$_{2}$ $\cdot$ H$_{2}$O} is found in reactions with q(Cu) = 0.25 and q(K) = 2 or 5 but not for q(K) = 1. Synthesis of this hydrated phase in high q(K) reinforces the idea exemplified in the q(Cu) = 1 reactions that varying q(K) can greatly change the amount of H$_{2}$O available in the flux to be incorporated into the product. Overall, we find that as q(K) increases, the amount of (OH)$^{-}$ and H$_{2}$O in the isolated product also increases. Accordingly, the only anhydrous oxide found formed with q(K) = 1. It is also important to note that increasing the amount of TeO$_{2}$ in solution by a factor of 4, to q(Cu) = 0.25, only altered the Cu:Te ratio in the resulting phase by a factor of 2. This, in combination with the study of the q(Cu) = 4 regime, implies that CuO is both easier to incorporate in the lattice and less soluble than TeO$_{2}$ in the hydroflux regime with q(K) = 1, 2, 5. This principle could direct the design of additional extended solids to target incorporation of particular Cu:Te ratios.

\section{Results and Discussion}
\subsection{K$_{2}$C\lowercase{u}$_{2}$T\lowercase{e}O$_{6}$ $\cdot$ H$_{2}$O}

\par
\begin{figure}[th]
    \centering
    \includegraphics[height = 1.3in, width = 3.2in]{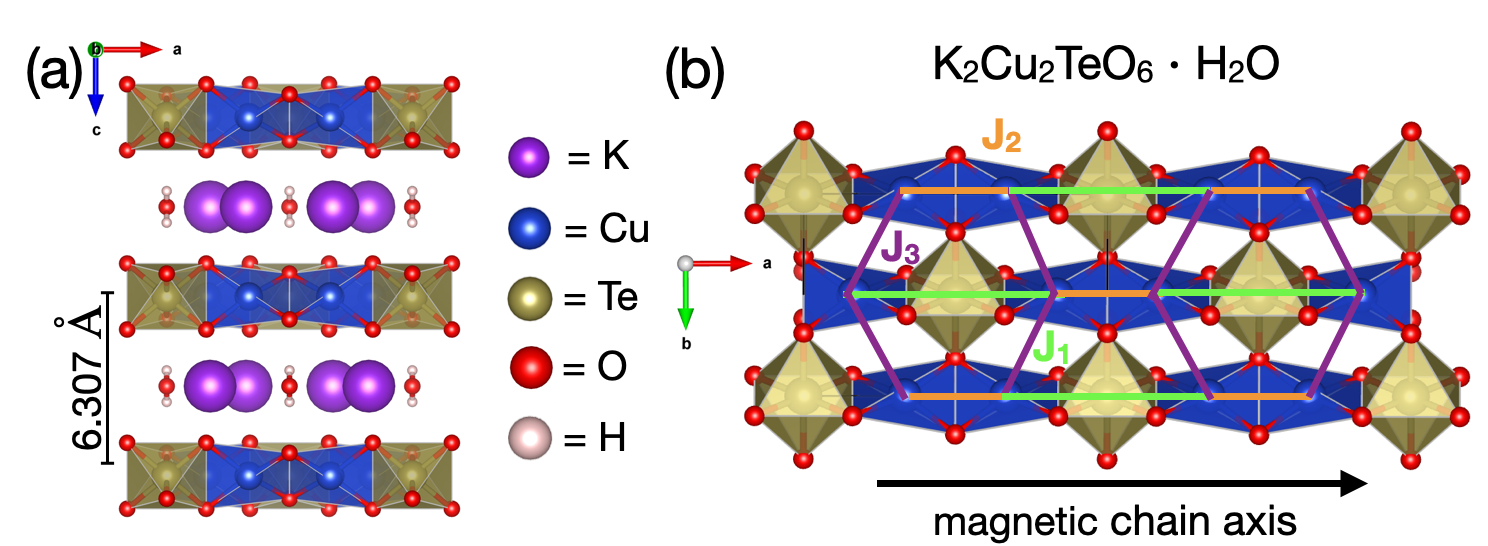}
    \caption{(a) Crystal structure of \mbox{K$_{2}$Cu$_{2}$TeO$_{6}$ $\cdot$ H$_{2}$O} showing the layered nature of the structure. (b) Crystal structure of \mbox{K$_{2}$Cu$_{2}$TeO$_{6}$ $\cdot$ H$_{2}$O} showing the 2D layers of square planar CuO$_{4}$ and octahedral TeO$_{6}$ in a honeycomb motif. The important magnetic interaction pathways between Cu sites on this motif are denoted as $J_{1}$, $J_{2}$, and $J_{3}$ following the convention of Na$_{2}$Cu$_{2}$TeO$_{6}$.\cite{Xu2005}}
    \label{fig:K2Cu2TeO6_H2O_structure}
\end{figure}

 Of the four K-Cu-Te-O(H) phases found in the hydroflux regime studied, \mbox{K$_{2}$Cu$_{2}$TeO$_{6}$ $\cdot$ H$_{2}$O} is the most simple structurally and will be the basis for our discussion of the magnetism found in these compounds. \mbox{K$_{2}$Cu$_{2}$TeO$_{6}$ $\cdot$ H$_{2}$O} is stabilized in a hydroflux reaction with q(K) = 2 and q(Cu) = 1 and can be synthesized phase pure as shown from pXRD in the SI. The structure of \mbox{K$_{2}$Cu$_{2}$TeO$_{6}$ $\cdot$ H$_{2}$O} was solved by SCXRD and the structure and single crystal data is shown in Fig. \ref{fig:K2Cu2TeO6_H2O_structure} and Table \ref{scxrd} respectively. In general, the structure is built from stacked layers of a CuO$_{4}$ honeycomb lattice with interpenetrating TeO$_{6}$ octahedra, with intercalated K$^{+}$ ions and H$_{2}$O molecules between the layers. One interesting structural feature in pXRD that is not captured in our single crystal data is the splitting of the $(00L)$ family of peaks. This indicates multiple species of \mbox{K$_{2}$Cu$_{2}$TeO$_{6}$ $\cdot$ H$_{2}$O} exist with a variation of interlayer spacing, likely due to different hydration levels. 

\par The magnetic properties of \mbox{K$_{2}$Cu$_{2}$TeO$_{6}$ $\cdot$ H$_{2}$O} are governed by its layered structure of distorted Cu$^{2+}$ spin-$\frac{1}{2}$ plaquettes and TeO$_{6}$ octahedra. In total, each Cu can be thought of as occupying a severely elongated octahedron as the apical Cu-O bond distances (omitted in Fig. \ref{fig:K2Cu2TeO6_H2O_structure}) are 2.606 \AA{} as is common in other Cu$^{2+}$ oxides.\cite{Xu2005} However, realistically, each Cu is bonded to four oxygen in a distorted square planar coordination with two short (1.990 \AA) and two longer (1.996 \AA) Cu-O bonds. Therefore, we discuss each CuO$_{6}$ octahedra as a CuO$_{4}$ square planar plaquette where each plaquette in the honeycomb has three distinct Cu nearest-neighbors shown in Fig. \ref{fig:K2Cu2TeO6_H2O_structure} by the labeled magnetic interactions. The closest neighbor is the discernible paired edge-sharing plaquette with a Cu-Cu distance of 2.89 \AA{} ($J_{2}$), then the apical-oxygen-sharing plaquette with a Cu-Cu distance of 3.25 \AA{} ($J_{3}$), and lastly a plaquette in an adjacent Cu-Cu edge-sharing pair with a Cu-Cu distance of 5.83 \AA{} ($J_{1}$). Additionally, there is a minimum distance of 4.028 \AA{} between planes along the c-axis with no magnetic interaction between layers.

\par Interestingly, \mbox{K$_{2}$Cu$_{2}$TeO$_{6}$ $\cdot$ H$_{2}$O} is isostructural to a well-studied system Na$_{2}$Cu$_{2}$TeO$_{6}$, except for the inclusion of interlayer H$_{2}$O units which are not present in the Na analogue.\cite{Xu2005} The Na analogue can be synthesized through a high-temperature solid state method but a similar method has never been reported for the K compound. Thorough study of the magnetism in Na$_{2}$Cu$_{2}$TeO$_{6}$ has enabled the characterization of the dominant $J_{1}$-$J_{2}$-$J_{3}$ interactions in this system.\cite{Xu2005, Koo2008, Sankar2014, Shangguan2021, Lin2022, Schmitt2014} In Na$_{2}$Cu$_{2}$TeO$_{6}$, the strongest interaction is $J_{1}$ which is an AFM interaction arising from super-superexchange along the Cu-O-O-Cu pathway. The remaining two interactions are both ferromagnetic (FM) with the strength of $J_{2}$ much larger than $J_{3}$ due to the geometry of the CuO$_{4}$ plaquettes and the Cu ${d_{x^{2}-y^{2}}}$ orbitals, resulting in very little orbital overlap along the interchain direction. As such, the magnetism of Na$_{2}$Cu$_{2}$TeO$_{6}$ is treated as an effective 1D system and we have taken a similar approach to understand the magnetism in \mbox{K$_{2}$Cu$_{2}$TeO$_{6}$ $\cdot$ H$_{2}$O}. 

\begin{figure}[th]
    \centering
    \includegraphics[height = 6in, width = 3.25in]{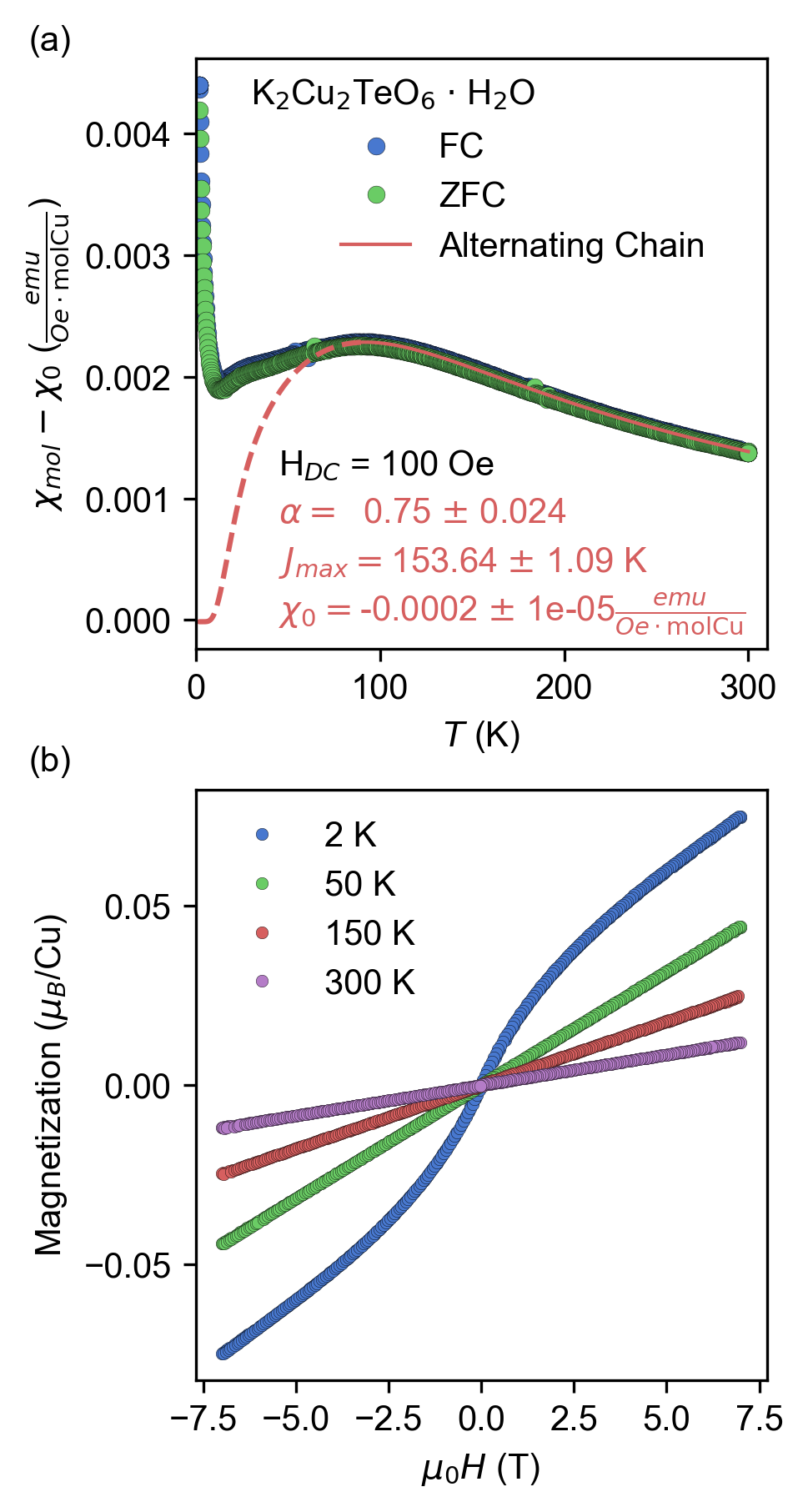}
    \caption{(a) Molar magnetic susceptibility of \mbox{K$_{2}$Cu$_{2}$TeO$_{6}$ $\cdot$ H$_{2}$O} powder measured using ZFC and FC protocols with an external field of H = 100 Oe showing low-dimensional behavior and no transitions indicative of long-range ordering. The high temperature data has been fit (red line) to a S = $\frac{1}{2}$ alternating exchange Heisenberg chain (discussed in main text). The upturn at low temperature is likely due to impurity paramagnetic phases or disorder. (b) Isothermal magnetization of \mbox{K$_{2}$Cu$_{2}$TeO$_{6}$ $\cdot$ H$_{2}$O} measured at T = 2, 50, 150, and 300 K in the range of $\mu_{0}H$ = $\pm$ 7 T applied external field.}

    \label{fig:K2Cu2TeO6_H2O_XvT_100Oe}
\end{figure}

The magnetic susceptibility of \mbox{K$_{2}$Cu$_{2}$TeO$_{6}$ $\cdot$ H$_{2}$O} shown in Fig. \ref{fig:K2Cu2TeO6_H2O_XvT_100Oe} displays clear 1D behavior and no distinct phase transitions down to T = 2 K. The large, broad hump which peaks at T $\approx$ 100 K is likely due to the onset of short-range correlations. A strong upturn at low temperatures is observed and is likely due to a small paramagnetic impurity phase or disorder in the system. Measurements on oriented single crystals of \mbox{K$_{2}$Cu$_{2}$TeO$_{6}$ $\cdot$ H$_{2}$O} would help resolve if this upturn is intrinsic to defects of the systems or due to an impurity phase from the hydroflux synthesis, but this measurement is currently difficult due to the small crystal size and agglomerated crystal habit. Isothermal magnetization measurements shown in Fig. \ref{fig:K2Cu2TeO6_H2O_XvT_100Oe}b show AFM behavior at all measured temperatures and the behavior at the T = 2 K isotherm dominated by the paramagnetic impurities in the system. In order to quantify and compare the magnetic behavior in this system, we have fit the magnetic susceptibility with a high-temperature series expansion (HTSE) for the alternating chain AFM Heisbenberg model.\cite{Johnston2000} In this system, assuming a much weaker $J_{3}$ term as is found in the Na analogue,\cite{Gao2020} the alternating chain Heisenberg model should accurately describe the magnetism as the $J_{1}$ and $J_{2}$ interactions together drive the magnetic interactions along the $a$-axis in this system. Using the HTSE derived in a previous work: \cite{Johnston2000} 

\begin{equation}
    \chi_{mol} = \chi_{0} + K\chi^{*}(\alpha, \dfrac{T}{J_{max}})
    \label{eqn:alt}
\end{equation}

the magnetic susceptibility from T = 90 - 300 K has been fit and using this model. For this model, $\chi_{0}$ describes any temperature-independent behavior, $K$ is a scaling term such that $K = \frac{N_{A}g^{2}\mu^{2}_{B}}{J_{max}k_{B}}$, $\alpha = \frac{J_{2}}{J_{1}}$, $J_{max} = J_{1}$, and $\chi^{*}$ is the modelled magnetic susceptibility. The $\alpha$ parameter is of particular importance because it delineates the two limits of the model: for $\alpha$ = 0 a simple spin dimer is recovered whereas for $\alpha$ = 1 a uniform AFM Heisenberg chain is recovered. 

When using the HTSE, a minimum temperature must be chosen for the fit range and, as such, all choices of minimum temperature and the corresponding extracted fit parameters are shown in the SI. For all fit ranges in which the errors in fit parameters remained reasonable, the average of each parameter and average of the error of each parameter have been reported; $\alpha$ = 0.75(2), $J_{max} = 153(1)$, and $K = 0.0139(1)$ are the results from this fitting procedure. Using this value of $K$ the Lande g-factor can be calculated as 2.38, close to the expected value of 2.2 for Cu$^{2+}$. The extracted value for $J_{max}$ is consistent with the overall shape of the magnetic susceptibility. Finally, an extracted value of $\alpha$ = 0.75 denotes the system as behaving between the dimer and uniform chain limits.

\par It is important to note that the application of the alternating chain model assumes both $J_{1}$ and $J_{2}$ are AFM in nature, so that $\alpha$ is strictly positive. This is not true for the Na analogue, where $J_{1}$ is AFM and $J_{2}$ is FM, but was initially a topic of controversy before neutron spectroscopy measurements were performed as the magnetic susceptibility could be modeled equally well with both alternating AFM-FM chain and alternating AFM-AFM chain models.\cite{Miura2006, Xu2005} Additionally, DFT results for the Na analogue had been reported supporting both the AFM and FM nature of the $J_{2}$ interaction.\cite{ Koo2008, Schmitt2014} As such, our approach of applying the HTSE for the alternating AFM-AFM chain only serves to distinguish the behavior in \mbox{K$_{2}$Cu$_{2}$TeO$_{6}$ $\cdot$ H$_{2}$O} from the two limits of the model, i.e. uniform chain and dimer behavior, as well as determine the magnitude of strongest magnetic interaction ($J_{max}$). With that said, application of the Goodenough-Kanamori rules indicates, qualitatively, a FM $J_{2}$ in \mbox{K$_{2}$Cu$_{2}$TeO$_{6}$ $\cdot$ H$_{2}$O}, similar to what has been determined in Na$_{2}$Cu$_{2}$TeO$_{6}$ and Na$_{3}$Cu$_{2}$SbO$_{6}$ by neutron spectroscopy.\cite{Gao2020, Shangguan2021, Miura2008}

\subsection{K$_{2}$C\lowercase{u}$_{2}$T\lowercase{e}O$_{6}$}

\begin{figure}[th]
    \centering
    \includegraphics[height = 1.665in, width = 3.2in]{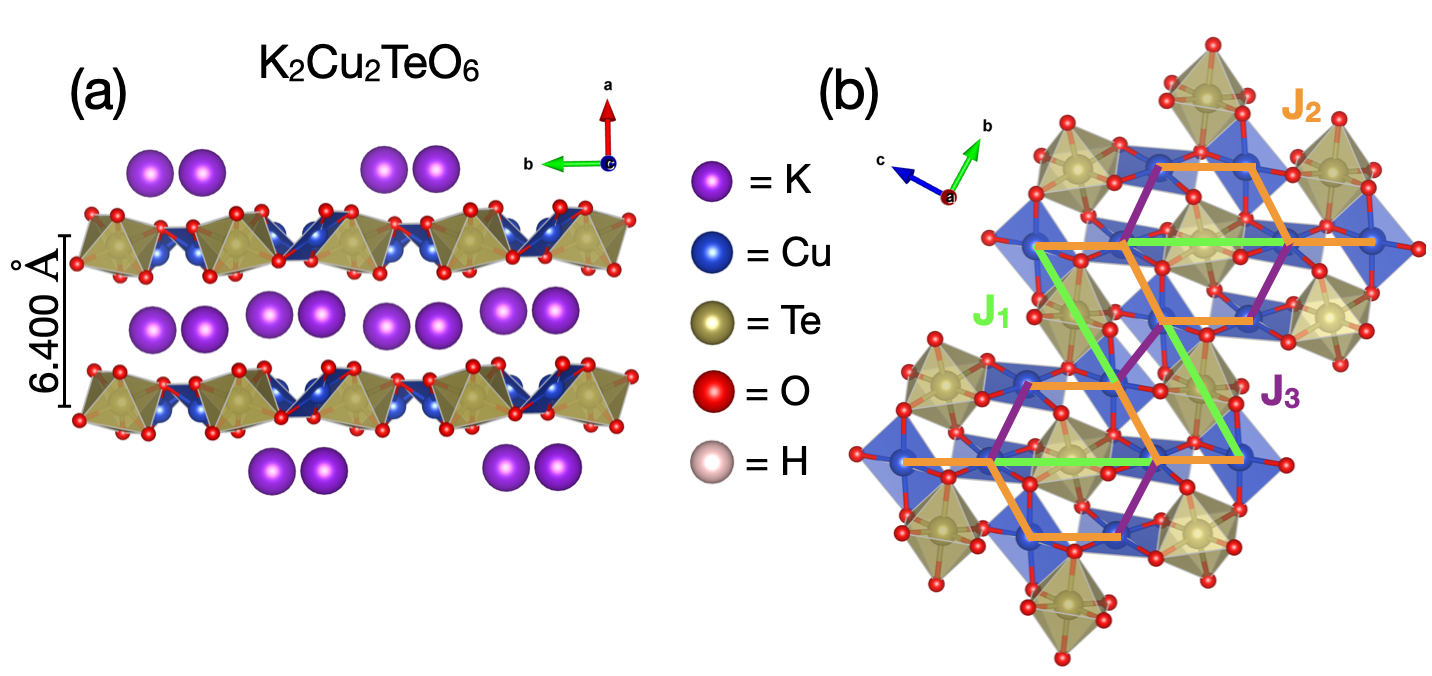}
    \caption{(a) Crystal structure of K$_{2}$Cu$_{2}$TeO$_{6}$ showing the layered nature of the structure. (b) Crystal structure of K$_{2}$Cu$_{2}$TeO$_{6}$ showing the 2D layers of square planar CuO$_{4}$ and octahedral TeO$_{6}$ in a honeycomb motif. The important magnetic interaction pathways on this motif are denoted as $J_{1}$, $J_{2}$, and $J_{3}$ following the convention used in this work.\cite{Xu2005}}
    \label{fig:K2Cu2TeO6_structure}
\end{figure}

K$_{2}$Cu$_{2}$TeO$_{6}$ was synthesized from the highest strength hydroflux tested, q(K) = 1, and was the only product to crystallize without incorporation of hydroxls or water into the structure. This suggests a fundamental shift in the chemistry of the hydroflux melt that prohibits hydrogen incorporation as water content in the flux is decreased to q(K) = 1 and the hydroflux becomes more oxidizing. Additional experimentation performed in a flux of hydrated hydroxide with no additional water did not yield novel phases. This phase can be made phase pure and does not readily hydrate with exposure to water and air to form \mbox{K$_{2}$Cu$_{2}$TeO$_{6}$ $\cdot$ H$_{2}$O} as seen in pXRD (SI) after washing the crystals with water to remove excess KOH from the samples. The structure of K$_{2}$Cu$_{2}$TeO$_{6}$ is shown in Fig. \ref{fig:K2Cu2TeO6_structure} and Table \ref{scxrd} and is very similar to the previously discussed \mbox{K$_{2}$Cu$_{2}$TeO$_{6}$ $\cdot$ H$_{2}$O} structure. 

\par As with \mbox{K$_{2}$Cu$_{2}$TeO$_{6}$ $\cdot$ H$_{2}$O}, the structure of K$_{2}$Cu$_{2}$TeO$_{6}$ is centered around the CuO$_{4}$ honeycomb lattice with interpenetrating TeO$_{6}$ octahedra. In this structure, Cu is bonded in a square planar coordination, with one short (1.970 \AA), two medium (1.983 \AA), and one long (2.058 \AA) planar Cu-O bonds. Apical Cu-O bonds of 2.496 \AA \text{} and 2.469 \AA \text{} justify treating the coordination as square planar rather than distorted octahedral. There are two long (3.108 \AA) and one short (3.058 \AA) Cu-Cu distances around the honeycomb lattice, $J_{2}$ and $J_{3}$, respectively, and interlayer K$^{+}$ ions segregate the Cu-Te-O layers.

\par The strength of each magnetic interaction in this phase can be understood using the basis derived from \mbox{K$_{2}$Cu$_{2}$TeO$_{6}$ $\cdot$ H$_{2}$O}; however in K$_{2}$Cu$_{2}$TeO$_{6}$, 1/3 of the CuO$_{4}$ plaquettes in each honeycomb are twisted $\approx$ 90$^{\circ}$ relative to the others and this seemingly minor structural difference drives stark changes in the magnetic behavior of the system.

\par In \mbox{K$_{2}$Cu$_{2}$TeO$_{6}$ $\cdot$ H$_{2}$O}, because the CuO$_{4}$ plaquettes all lie roughly in the same plane, the magnetic behavior is best described as an alternating chain along the $a$ direction, and plaquettes  effectively do not couple via interactions along the $b$ and $c$ directions. In K$_{2}$Cu$_{2}$TeO$_{6}$, the tilting of 1/3 of these CuO$_{4}$ plaquettes by $\approx$ 90$^{\circ}$ changes the Cu-Cu interactions to become more two-dimensional as both $J_{1}$ and $J_{2}$ are no longer fixed to one direction (shown in Fig. \ref{fig:K2Cu2TeO6_structure}) within the honeycomb plane. 

\par Specifically, this two-dimensional behavior arises due to each CuO$_{4}$ plaquette no longer sharing an edge with another plaquette in a pair, but instead sharing corners with two plaquettes along different directions. This leads to altered strengths and directions of magnetic interactions $J_{1}$ and $J_{2}$. Although $J_{1}$ gains dimensionality in K$_{2}$Cu$_{2}$TeO$_{6}$, the strength of the interaction likely remains the same, as it remains a super-superexchange Cu-O-O-Cu AFM interaction. The FM $J_{2}$ interaction, which we suppose is significantly weaker than $J_{1}$ in \mbox{K$_{2}$Cu$_{2}$TeO$_{6}$ $\cdot$ H$_{2}$O} based on neutron spectroscopy studies of isostructural Na$_{2}$Cu$_{2}$TeO$_{6}$,\cite{Gao2020, Shangguan2021} is likely even weaker in K$_{2}$Cu$_{2}$TeO$_{6}$. The edge-sharing plaquettes of \mbox{K$_{2}$Cu$_{2}$TeO$_{6}$ $\cdot$ H$_{2}$O} provide two oxygen avenues for Cu-O-Cu superexchange within a pair of plaquettes, but the corner sharing plaquettes of K$_{2}$Cu$_{2}$TeO$_{6}$ have only one pathway for Cu-O-Cu superexchange for a given Cu-Cu pair. This $J_{2}$ interaction now describes the ability of the CuO$_{4}$ plaquette to singly couple equally to two neighboring plaquettes within the $bc$-plane, rather than doubly couple to a single neighboring plaquette along one axis. Finally, $J_{3}$ remains effectively unchanged but now describes one interaction along the $b$-axis between adjacent CuO$_{4}$ plaquettes which do not share a face or an edge. This interaction occurs once per CuO$_{4}$ plaquette and twice within a given hexagon because each dimer pair in K$_{2}$Cu$_{2}$TeO$_{6}$ interacts with an adjacent pair through a $J_{2}$ interaction rather than exclusively a $J_{3}$ interaction. These considerations indicate that the magnetic behavior in K$_{2}$Cu$_{2}$TeO$_{6}$ is likely to be more two-dimensional.

\begin{figure}[t]
    \centering
    \includegraphics[height = 6in, width = 3.25in]{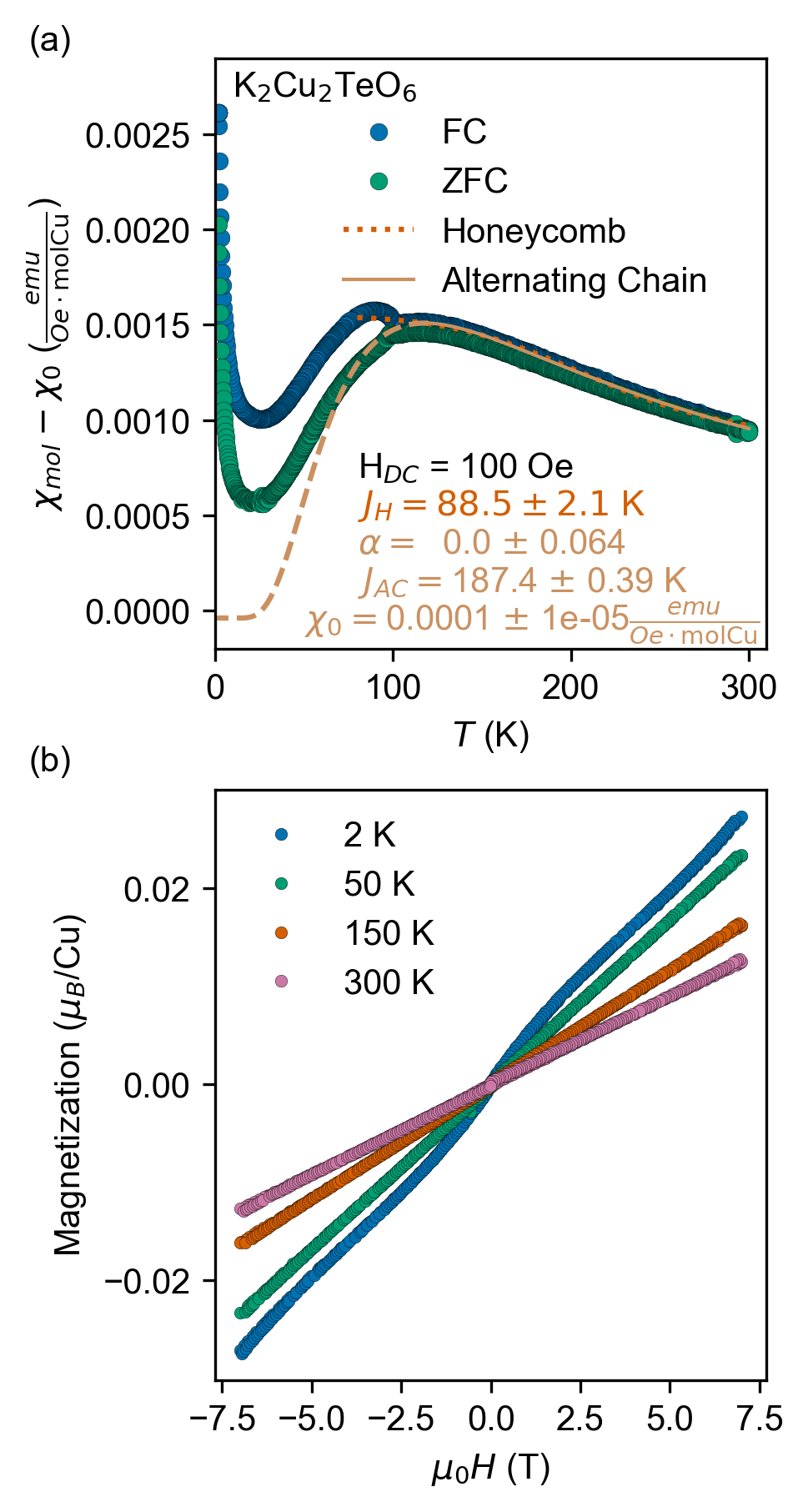}
    \caption{(a) Molar magnetic susceptibility of K$_{2}$Cu$_{2}$TeO$_{6}$ powder measured using ZFC and FC protocols with an external field of H = 100 Oe showing low-dimensional behavior and long-range magnetic ordering at T$_{N}$ = 100 K indicative of antiferromagnetic behavior. The high temperature data has been fit to a nearest-neighbor honeycomb model (red) and to a S = $\frac{1}{2}$ alternating exchange Heisenberg chain (purple). The upturn at low temperature is likely due to impurity paramagnetic phases or disorder in K$_{2}$Cu$_{2}$TeO$_{6}$. (b) Isothermal magnetization of K$_{2}$Cu$_{2}$TeO$_{6}$ measured at T = 2, 50, 150, and 300 K in the range of $\mu_{0}H$ = $\pm$ 7 T applied external field.}
    \label{fig:K2Cu2TeO6_XvT_100Oe}
\end{figure}

Despite the small structural differences between K$_{2}$Cu$_{2}$TeO$_{6}$ and its hydrated analogue, a clear difference in the magnetic behavior is observed. As shown in Fig. \ref{fig:K2Cu2TeO6_XvT_100Oe}, K$_{2}$Cu$_{2}$TeO$_{6}$ shows the onset of AFM order at T$_{N}$ = 100 K, while \mbox{K$_{2}$Cu$_{2}$TeO$_{6}$ $\cdot$ H$_{2}$O} shows no order down to T = 2 K. The onset of long-range magnetic order is likely due to the previously discussed effects of the structural changes on the geometry of the exchange interactions. Additionally, the magnetic susceptibility shows a sharp upturn at low temperatures due to either small amounts of paramagnetic impurity or intrinsic disorder in the system. Isothermal magnetization measurements shown in Fig. \ref{fig:K2Cu2TeO6_XvT_100Oe} demonstrate the change in behavior below T$_{N}$ = 100 K and the low effective moment in this system. In order to quantify deviation from alternating chain AFM behavior, we fit the high temperature magnetic susceptibility to two models. One model is the alternating AFM chain,\cite{Johnston2000} which in this compound propagates in two directions in the $bc$-plane as $J_{1}$ and $J_{2}$ are no longer constrained to one dimension, and the other model is the nearest-neighbor honeycomb model,\cite{Rushbrooke1958} which assumes the $J_{2}$ interaction is both consistent around the honeycomb and is larger in magnitude than the super-superexchange $J_{1}$ interaction. We consider these two models as two ends of a spectrum of possible behavior in this system and extract $J$ values corresponding to each model to determine general agreement. The model used for the alternating AFM chain is Equation \ref{eqn:alt} and the Rushbrook-Woods model\cite{Rushbrooke1958} was used to fit the nearest-neighbor honeycomb model:

\begin{equation}
    \chi_{mol} = \chi_{0} + \left(\frac{K}{T} \right)\frac{S(S+1)}{1 + Ax + Bx^2 + Cx^3 + Dx^4 + Ex^5 + Fx^6}    
    \label{eqn:honeycomb}
\end{equation}

where $S = \frac{1}{2}$ for a Cu$^{2+}$ system, $K$ is an overall scaling factor including fundamental constants, $x = J/k_{B}T$, and the constants $A-F$ are derived in the aforementioned work.\cite{Rushbrooke1958} Using the HTSE for the alternating AFM chain Heisenberg model, we fit the magnetic susceptibility data from T = 100 - 300 K with all minimum temperatures for which the model and extracted parameters remained stable, as shown in the SI. From the averages of these fittings, we extracted $J_{max}$ = 187 K and $\alpha$ = 0. This extracted $J_{max}$ is in agreement with the $J_{1}$ value obtained via our HTSE fit of \mbox{K$_{2}$Cu$_{2}$TeO$_{6}$ $\cdot$ H$_{2}$O} in this work and what was determined for the Na analogue.\cite{Gao2020, Xu2005} This is consistent with our structural analysis of $J_{1}$ having the same geometry and likely same sign in each compound. As discussed previously, $\alpha$ = 0 suggests the dimer limit of the alternating chain model. This dimer-like behavior likely develops in K$_{2}$Cu$_{2}$TeO$_{6}$ as compared to \mbox{K$_{2}$Cu$_{2}$TeO$_{6}$ $\cdot$ H$_{2}$O} due the weakening of $J_{2}$ in overall magnitude and directional magnitude caused by its propagation in multiple directions of the $bc$-plane through interactions between corner-sharing plaquettes.

On the other end of the spectrum, we fit the magnetic susceptibility data from T = 100 - 300 K to a HTSE of the nearest-neighbor honeycomb model, which considers $J_{2}$ as the sole interaction. Note that this model assumes $J_{2}$ is a consistent interaction throughout the ring, while we have determined based on the structure of K$_{2}$Cu$_{2}$TeO$_{6}$ that a different weak interaction, $J_{3}$, likely occurs between non-corner-sharing CuO$_{4}$ plaquettes. The value of $J_{2}$ = 88.5 K extracted from this model is on the same order as the $J_{2}$ = 8.7 meV ($\approx$ 100 K) determined from neutron measurements on Na$_{2}$Cu$_{2}$TeO$_{6}$.\cite{Gao2020, Shangguan2021} As both the honeycomb and alternating AFM chain models fit the high temperature data well, the magnetic behavior is likely described by some combination of these two at the microscopic level. As previously discussed, the two-dimensional magnetic interactions in this phase are likely the key factor that allow for long range AFM ordering. A naive tiling of the Cu$^{2+}$ moments with three adjacent spins up and three adjacent spins down within a hexagon can satisfy the expected AFM $J_{1}$ and FM $J_{2}$ interactions; this tiling results in a $k = (0,0,0)$ magnetic propagation vector (shown in SI). Neutron diffraction measurements to definitively characterize the magnetic ordering in this system are underway and will be fundamental to understanding how these small changes in the honeycomb lattice structure lead to distinct magnetic behaviors.

\subsection{K$_{6}$C\lowercase{u}$_{9}$T\lowercase{e}$_{4}$O$_{24}$ $\cdot$ 2 H$_{2}$O}

\begin{figure}[th]
    \centering
    \includegraphics[height = 1.35in, width = 3in]{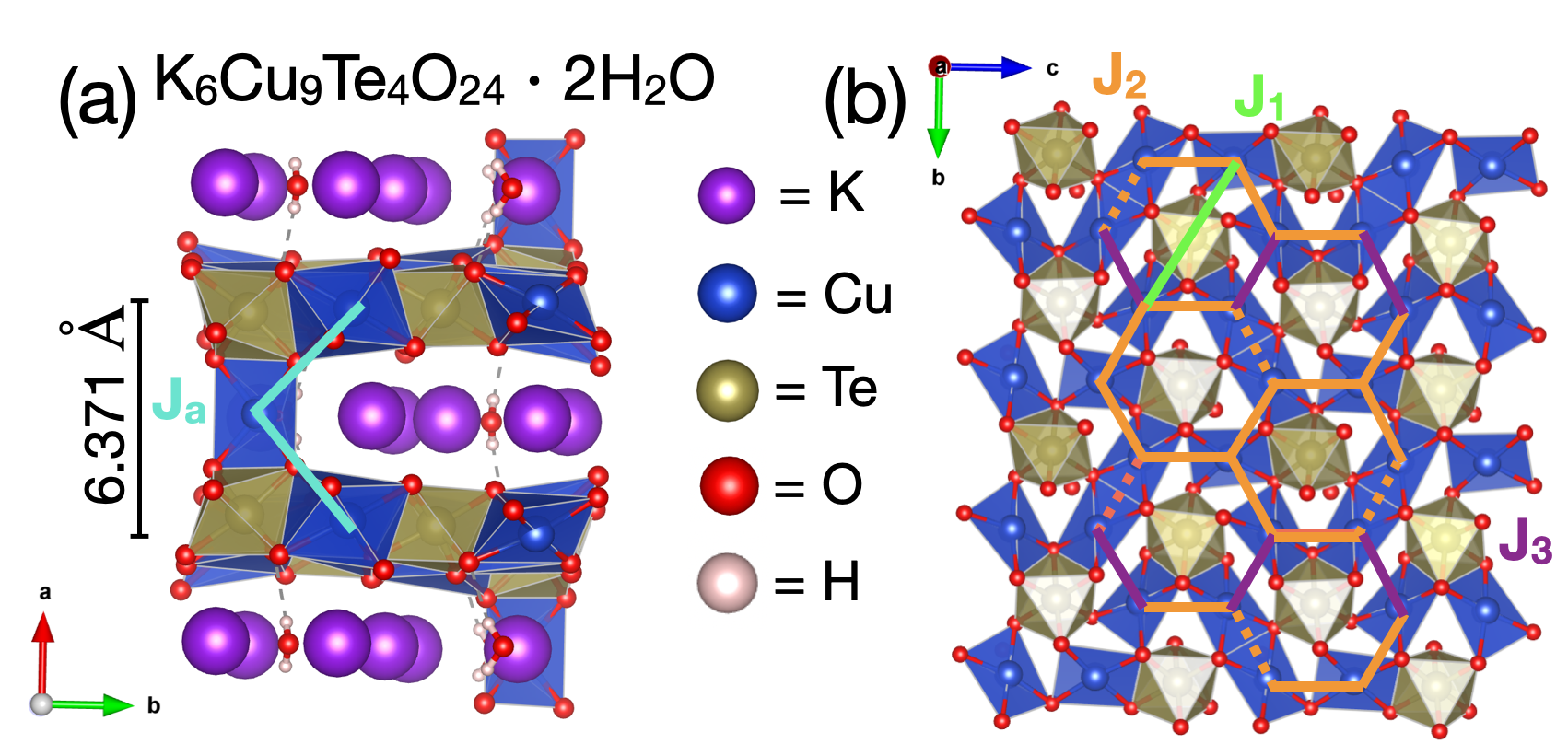}
    \caption{(a) Crystal structure of \mbox{K$_{6}$Cu$_{9}$Te$_{4}$O$_{24}$ $\cdot$ 2 H$_{2}$O} showing the layered nature of the structure.  $J_{a}$ denotes the interlayer coupling present in \mbox{K$_{6}$Cu$_{9}$Te$_{4}$O$_{24}$ $\cdot$ 2 H$_{2}$O} caused by CuO$_{4}$ plaquettes that bridge the Cu-Te honeycomb layers. (b) Crystal structure of \mbox{K$_{6}$Cu$_{9}$Te$_{4}$O$_{24}$ $\cdot$ 2 H$_{2}$O} showing the 2D layers of square planar CuO$_{4}$ and octahedral TeO$_{6}$ in a honeycomb motif. The important magnetic interaction pathways on this motif are denoted as $J_{1}$, $J_{2}$, $J_{3}$ following the convention used in this work.\cite{Xu2005} The dotted $J_{2}$ interaction is used to denote the different but similar $J_{2}$ interactions possible (discussed in the text). The bridging plaquettes responsible for $J_{a}$ occur in the rings without the $J_{3}$ interactions. Note that this phase belongs to a polar space group.}
    \label{fig:K6Cu9Te4O24_2H2O_structure}
\end{figure}

\mbox{K$_{6}$Cu$_{9}$Te$_{4}$O$_{24}$ $\cdot$ 2 H$_{2}$O} was synthesized from a dilute hydroflux of q(K) = 5 and contains water molecules within its unit cell, following the trend of high water content within the hydroflux resulting in hydroxl and water incorporation into crystallized structures. The structure is shown in Fig. \ref{fig:K6Cu9Te4O24_2H2O_structure} and Table \ref{scxrd} and can be synthesized phase pure, as shown by pXRD in the SI. Like the other phases synthesized with q(Cu) = 1, \mbox{K$_{6}$Cu$_{9}$Te$_{4}$O$_{24}$ $\cdot$ 2 H$_{2}$O} exhibits a layered structure of CuO$_{4}$ and TeO$_{6}$ tiling a structural honeycomb motif, as similarly distorted CuO$_{6}$ octahedra yield effective CuO$_{4}$ square planar plaquettes. However, unlike the other phases synthesized in this study, \mbox{K$_{6}$Cu$_{9}$Te$_{4}$O$_{24}$ $\cdot$ 2 H$_{2}$O} contains a CuO$_{4}$ plaquette which bridges the honeycomb layers, resulting in a distinctly three-dimensional structure and a higher ratio of Cu:Te. There is a high degree of distortion in this structure, with five distinct Cu sites. While each Cu remains in an effective square planar coordination to oxygen, the apical Cu-O bond distances vary significantly. For the in-plane Cu ions, \mbox{K$_{6}$Cu$_{9}$Te$_{4}$O$_{24}$ $\cdot$ 2 H$_{2}$O} shows the smallest apical Cu-O bond of all q(Cu) = 1 phases in this work, with apical Cu-O bond lengths ranging from 2.302 \AA \text{} to 2.787 \AA \text{} and significant asymmetry in apical distances within each octahedron. For the interplanar Cu, only one apical Cu-O bond is possible, resulting in a distorted square pyramidal geometry. Compared to the other phases, each with one distinct Cu site and nearly symmetric apical bond lengths which vary by no more than 0.03 \AA, this phase is anomalous and heavily distorted. Furthermore, some CuO$_{4}$ plaquettes are in-plane, like in \mbox{K$_{2}$Cu$_{2}$TeO$_{6}$ $\cdot$ H$_{2}$O} and some are twisted 90$^{\circ}$ relative to neighboring plaquettes, like in K$_{2}$Cu$_{2}$TeO$_{6}$ increasing the complexity of the structure. 

Due to the large degree of distortion in \mbox{K$_{6}$Cu$_{9}$Te$_{4}$O$_{24}$ $\cdot$ 2 H$_{2}$O}  resulting in five distinct Cu sites, several short apical Cu-O bond distances, and an added dimensionality due to the interplanar CuO$_{4}$ plaquette, the potential magnetic interactions are many and complex. Two inequivalent honeycomb rings, one with a CuO$_{4}$ interplanar linker, and one without, are required to describe the magnetic interactions and are shown in Fig. \ref{fig:K6Cu9Te4O24_2H2O_structure}. The ring containing the CuO$_{4}$ interplanar linker can be described with a combination of $J_{1}$, $J_{2}$, and $J_{a}$ interactions. The ring that does not contain the CuO$_{4}$ interplanar linker can be described with a combination of $J_{1}$, $J_{2}$, and $J_{3}$ interactions. As in the other phases synthesized in this work, $J_{1}$ describes the super-superexchange AFM Cu-O-O-Cu interaction across the honeycomb, which due to the inequivalent Cu sites can occur along three inequivalent pathways. If the pathway lengths and angles were equivalent, the three $J_{1}$ interactions across the honeycomb would cancel each other out; because the pathways are inequivalent, the overall effect of the $J_{1}$ interactions is likely non-zero but greatly reduced compared to other phases in this work. $J_{2}$ describes the FM coupling between CuO$_{4}$ plaquettes, which in this phase can occur either doubly between edge-sharing plaquettes along two slightly inequivalent pathways (dotted line in Fig. \ref{fig:K6Cu9Te4O24_2H2O_structure}), or singly between corner-sharing plaquettes. $J_{3}$ describes the weak interaction between plaquettes which share neither a corner nor an edge, requiring the participation of distant apical oxygens. $J_{a}$ describes the likely weak interaction between in-layer plaquettes and the bridging plaquette along the $a$-axis which introduces the non-negligible three-dimensionality to the magnetic interactions in this system. This interaction occurs in four inequivalent ways and also requires the participation of apical oxygens, so it is likely to be relatively weak due to the elongated apical Cu-O bonds.

\begin{figure}[t]
    \centering
    \includegraphics[height = 6in, width = 3.25in]{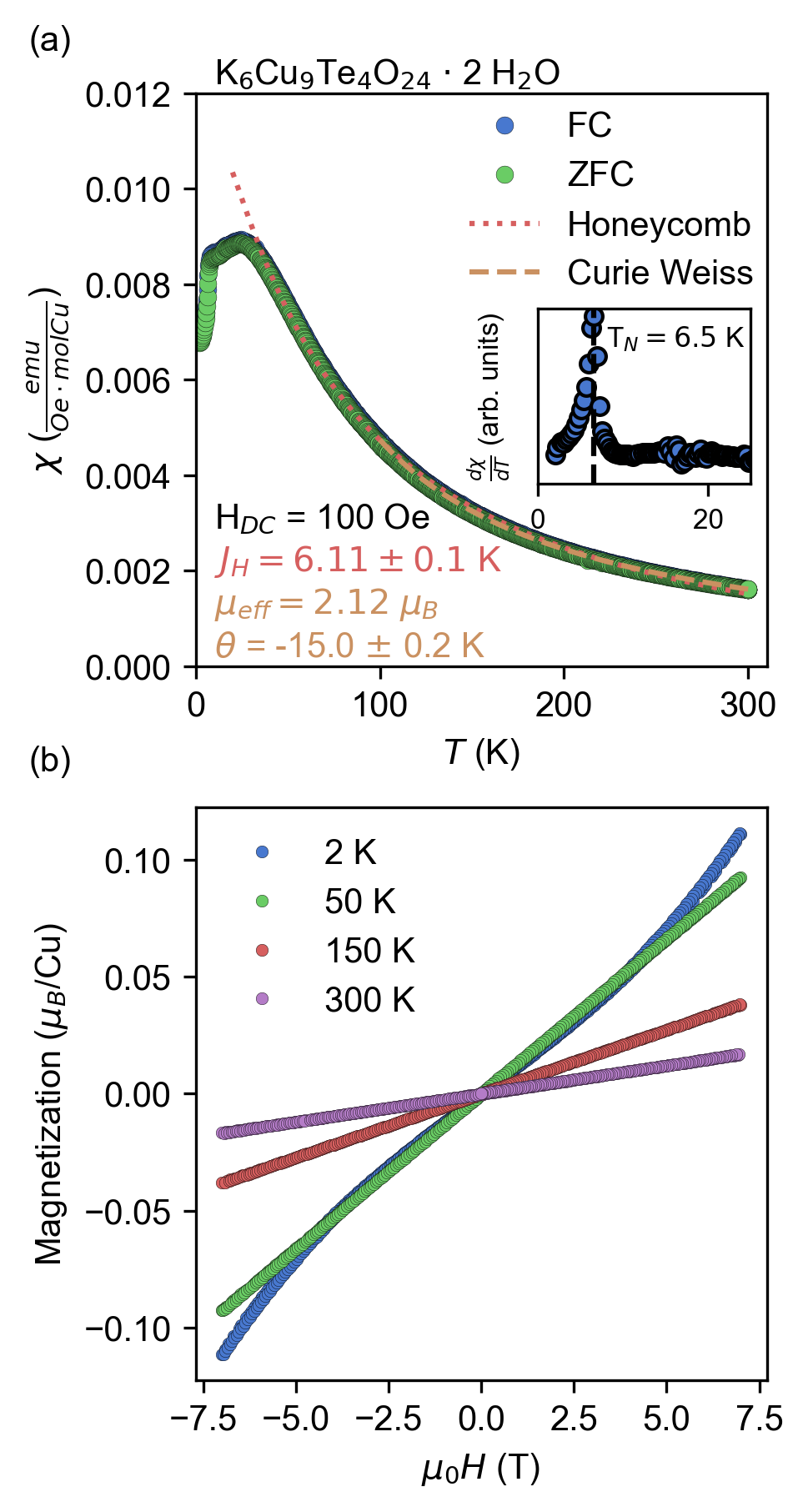}
    \caption{Molar magnetic susceptibility of \mbox{K$_{6}$Cu$_{9}$Te$_{4}$O$_{24}$ $\cdot$ 2 H$_{2}$O} powder measured using ZFC and FC protocols with an external field of H = 100 Oe showing Curie-Weiss behavior at high temperatures and antiferromagnetic ordering at T$_{N}$ = 6.5 K revealed by analyzing the derivative of the magnetic susceptibility with respect to temperature (inset). The high temperature data has been fit to a nearest-neighbor honeycomb model (red) and to Curie-Weiss behavior (purple) and the extracted values from these fits are shown. (b) Isothermal magnetization of \mbox{K$_{6}$Cu$_{9}$Te$_{4}$O$_{24}$ $\cdot$ 2 H$_{2}$O} measured at T = 2, 50, 150, and 300 K in the range of $\mu_{0}H$ = $\pm$ 7 T applied external field.}
    \label{fig:K6Cu9Te4O24_XvT_100Oe}
\end{figure}

The magnetic susceptibility of \mbox{K$_{6}$Cu$_{9}$Te$_{4}$O$_{24}$ $\cdot$ 2 H$_{2}$O} shown in Fig. \ref{fig:K6Cu9Te4O24_XvT_100Oe} shows Curie-Weiss behavior before the onset of long range magnetic AFM order at T$_{N}$ = 6.5 K (emphasized by the inset of Fig. \ref{fig:K6Cu9Te4O24_XvT_100Oe}). This AFM order at T$_{N}$ = 6.5 K is supported by the isothermal magnetization measurements in Fig. \ref{fig:K6Cu9Te4O24_XvT_100Oe} with the T = 2 K isotherm showing markedly different behavior. Additionally, the T = 2 K isotherm appears to begin an upturn above $\mu_{0}H$ = 6 T indicating the system may exhibit metamagnetic behavior but higher field and single crystal measurements are needed to resolve this anomalous behavior. In order to quantify the magnetic behavior of this phase, we have fit the magnetic susceptibility in the high temperature regime with two models. First, as our geometric considerations showed the importance of the $J_{2}$ interactions, we performed a fitting using the nearest-neighbor honeycomb model using the Rushbrook-Woods model\cite{Rushbrooke1958} and Equation \ref{eqn:honeycomb}. Using the nearest-neighbor honeycomb model, we extracted $J$ = 6.11 K, which is in close agreement with the observed ordering temperature. We also performed a Curie-Weiss fitting:

\begin{equation}
    \chi_{mol} = \chi_{0} + \dfrac{C}{T - \Theta}   
    \label{eqn:CW}
\end{equation}

where $C$ is the Curie constant which can be converted to effective magnetic moment ($\mu_{eff} = \sqrt{(8C)}$) and $\Theta$ is the Weiss temperature, which gives an indication of predominant interactions of the system. Unlike the other compounds synthesized in this work, the high temperature behavior of \mbox{K$_{6}$Cu$_{9}$Te$_{4}$O$_{24}$ $\cdot$ 2 H$_{2}$O} is Curie-Weiss-like. Using this model, we extracted $\mu_{eff}$ = 2.12 $\mu_{B}$, which is larger than the expected value of 1.73 $\mu_{B}$ for an isolated $S = \frac{1}{2}$ system due to the presence of spin-orbit coupling, and  $\Theta$ = -15.0 K which indicates antiferromagnetic behavior and is close to the observed T$_{N}$ = 6.5 K AFM ordering. Additionally, applying the mean-field expression for the Weiss temperature, we can extract the effective $J_{CW}$ via: $\Theta = \frac{-zJ_{CW}S(S+1)}{3k_B}$ where $z$ is the number of coordinate spins (3 for a honeycomb) and $S = \frac{1}{2}$ - this yields $J_{CW}$ = 20 K which is very close to our extracted $J$ from our HTSE honeycomb fitting as well as the experimentally observed AFM ordering temperature.

\par The strong agreement with the Curie-Weiss law above the ordering temperature suggests that the structural three-dimensionality and overall weak magnetic interactions of this system plays a large role in its magnetic ordering. With the complex structure and many inequivalent magnetic interactions, \mbox{K$_{6}$Cu$_{9}$Te$_{4}$O$_{24}$ $\cdot$ 2 H$_{2}$O} is far from a model system and development of an effective Hamiltonian to describe the magnetic order will be difficult. Additional studies are required to elucidate the nature and cause of the AFM ordering. This system also belongs to a polar space group, and the interaction of this polarity with the magnetism in this compound should be explored further.

\subsection{K$_{2}$C\lowercase{u}T\lowercase{e}O$_{4}$(OH)$_{2}$ $\cdot$ H$_{2}$O}

\par
\begin{figure}[th]
    \centering
    \includegraphics[height = 1.665in, width = 3.2in]{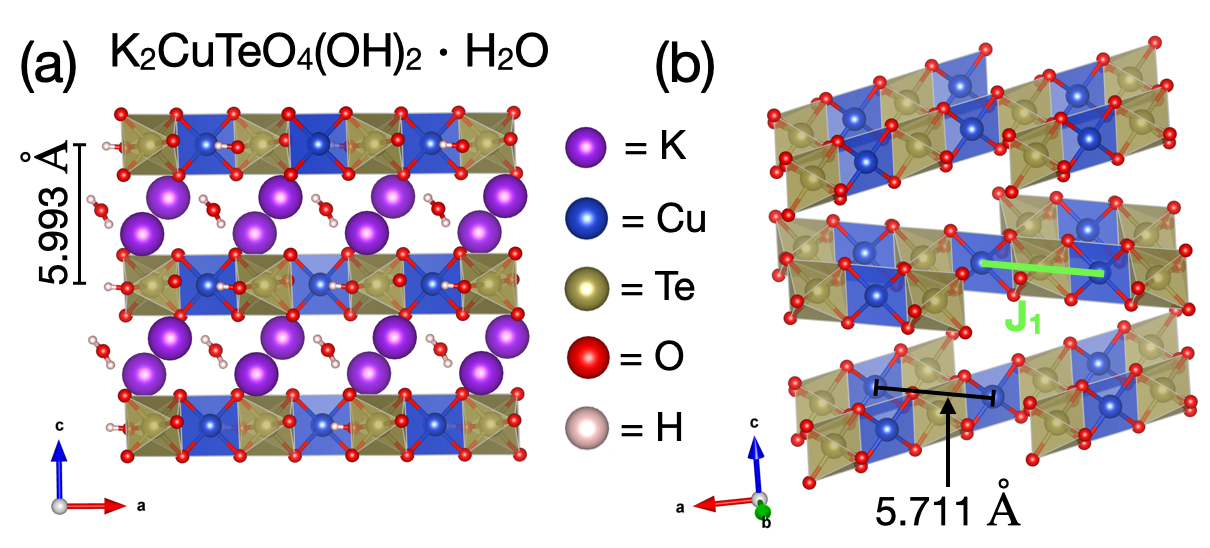}
    \caption{(a) Crystal structure of \mbox{K$_{2}$CuTeO$_{4}$(OH)$_{2}$ $\cdot$ H$_{2}$O} showing the layered nature of the structure. (b) Crystal structure of \mbox{K$_{2}$CuTeO$_{4}$(OH)$_{2}$ $\cdot$ H$_{2}$O} showing the 2D layers of square planar CuO$_{4}$ and octahedral TeO$_{6}$ in a 1D chain motif (interstitial K and H$_{2}$O have been omitted here for clarity). The important magnetic interaction pathway in this motif are denoted as $J_{1}$. The closest interlayer and intralayer Cu-Cu distances between chains are denoted in (a) and (b) respectively.}
    \label{fig:K2CuTeO4(OH)2_2H2O_structure}
\end{figure}

 As previously discussed, all the compounds in the q(Cu) = 1 regime formed layered structures containing a honeycomb motif of Cu and Te octahedra, with an approximately 2:1 ratio of Cu:Te. In hopes to push the formation of different magnetic motifs containing Cu$^{2+}$ species, we limited the ratio of Cu and Te in solution. \mbox{K$_{2}$CuTeO$_{4}$(OH)$_{2}$ $\cdot$ H$_{2}$O} was the only non-binary phase formed in the q(Cu) = 0.25 regime regardless of the q(K) ratio used. The structure has been previously reported \cite{Effenberger1993} and the structure solved by SCXRD is in agreement with this previously reported solution (shown in Fig. \ref{fig:K2CuTeO4(OH)2_2H2O_structure} and Table \ref{scxrd}). pXRD (SI) shows that the product can be synthesized phase pure. Previous reports indicate \mbox{K$_{2}$CuTeO$_{4}$(OH)$_{2}$ $\cdot$ H$_{2}$O} was synthesized under "moderate" hydrothermal conditions but the circumstances of its formation had not been understood as it was previously the only known phase in the K-Cu-Te-O(H) phase space. To put this into context, Fig. \ref{fig:K2CuTeO4(OH)2_2H2O_structure} shows the structure of \mbox{K$_{2}$CuTeO$_{4}$(OH)$_{2}$ $\cdot$ H$_{2}$O} demonstrating that the decreased q(Cu) = 0.25 pushes the structure away from the honeycomb motif present when q(Cu) = 1. Additionally, for q(Cu) = 1 the resultant structures maintains a ratio of Cu:Te $\approx$ 2, whereas; in \mbox{K$_{2}$CuTeO$_{4}$(OH)$_{2}$ $\cdot$ H$_{2}$O} the ratio of Cu:Te = 1 even though the q(Cu) value is decreased by a factor of 4. Thus, we can say that in this regime, the incorporation of Te into the structure is the limiting factor in these syntheses. This is especially true considering the previously discussed observation that for all reactions with q(Cu) = 4, the primary product was consistently only CuO.

\par Beyond the successful synthesis of \mbox{K$_{2}$CuTeO$_{4}$(OH)$_{2}$ $\cdot$ H$_{2}$O} in a hydroflux regime, the resulting magnetic motif is of particular interest. The crystal structure shown in Fig. \ref{fig:K2CuTeO4(OH)2_2H2O_structure} demonstrates that due to the reduced ratio of Cu:Te, the 2D honeycomb motif present in the other systems has broken apart. In \mbox{K$_{2}$CuTeO$_{4}$(OH)$_{2}$ $\cdot$ H$_{2}$O}, planes of chains of alternating CuO$_{4}$ square planar plaquettes and TeO$_{4}$(OH)$_{2}$ octahedra are stacked along the $c$-axis and there exist no magnetic pathways between these chains in the $ab$-plane. Each CuO$_{4}$ square planar plaquette has square planar Cu-O bond lengths of 1.96 \AA{} and apical Cu-O bond distances of 3.28 \AA{}. This distortion is much larger than the other compounds discussed in this work and highlights the structural 1D character of \mbox{K$_{2}$CuTeO$_{4}$(OH)$_{2}$ $\cdot$ H$_{2}$O}. An additional important feature of \mbox{K$_{2}$CuTeO$_{4}$(OH)$_{2}$ $\cdot$ H$_{2}$O} is the replacement of oxygen with (OH)$^{-}$ ligands in the Te octahedra. This is likely a result of the low hydroxide concentration used to synthesize this compound, i.e. \mbox{K$_{2}$CuTeO$_{4}$(OH)$_{2}$ $\cdot$ H$_{2}$O} was formed for q(K) = 2 and 5 but not for q(K) = 1.

\mbox{K$_{2}$CuTeO$_{4}$(OH)$_{2}$ $\cdot$ H$_{2}$O} is a truly structurally and magnetically 1D motif whereby the only readily apparent magnetic interaction is a nearest-neighbor intrachain interaction $J_{1}$. As in the other reported systems, this $J_{1}$ involves a Cu-O-O-Cu super-superexchange pathway. These 1D chains form layers with the $ac$-plane but there is no bonding between these chains within the layers, and a large intralayer chain distance (5.771 \AA) is observed. Additionally, these layers of isolated chains are separated by interlayer K$^{+}$ ions and H$_{2}$O molecules along the $b$-axis. Importantly, the interlayer chain separation is only slightly larger (5.993 \AA) than the intralayer chain separation. While the large distances in both cases suggest they would host weak interchain interactions, the presence of more substantial bonding along the interlayer direction suggests coupling is more likely to occur out of plane.  Based on this geometric analysis, \mbox{K$_{2}$CuTeO$_{4}$(OH)$_{2}$ $\cdot$ H$_{2}$O} stands out as a potential candidate for the study of the nearest-neighbor 1D Heisenberg chain. 

\begin{figure}[th]
    \centering
    \includegraphics[height = 6in, width = 3.25in]{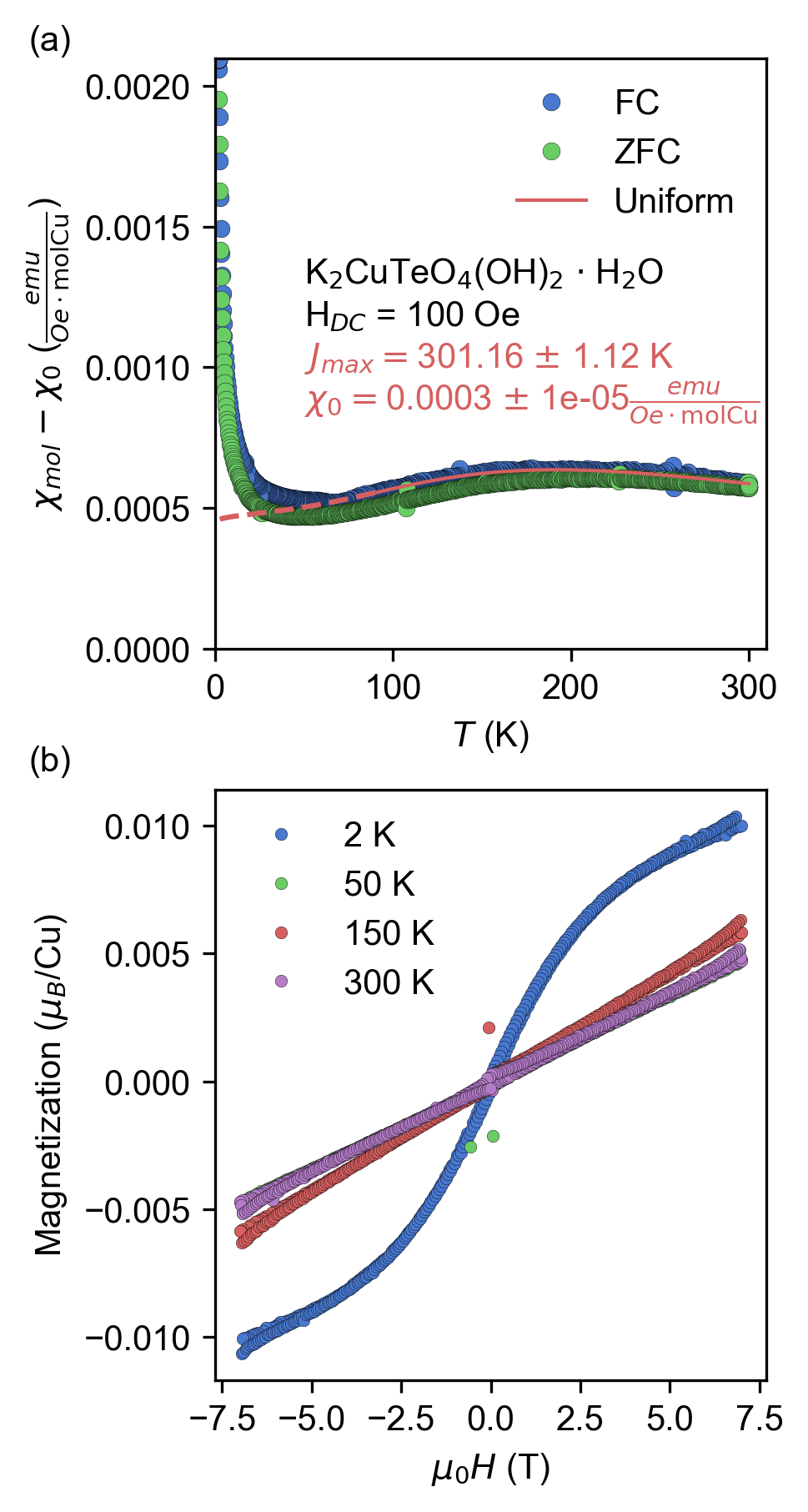}
    \caption{Molar magnetic susceptibility of \mbox{K$_{2}$CuTeO$_{4}$(OH)$_{2}$ $\cdot$ H$_{2}$O} powder measured using ZFC and FC protocols with an external field of H = 100 Oe showing low-dimensional behavior and no transitions indicative of long-range ordering. The high temperature data has been fit (red line) to a S = $\frac{1}{2}$ uniform exchange Heisenberg chain (discussed in main text). The upturn at low temperature and small deviation between ZFC and FC curves are likely due to impurity paramagnetic phases or disorder in \mbox{K$_{2}$CuTeO$_{4}$(OH)$_{2}$ $\cdot$ H$_{2}$O}. (b) Isothermal magnetization of \mbox{K$_{2}$CuTeO$_{4}$(OH)$_{2}$ $\cdot$ H$_{2}$O} measured at T = 2, 50, 150, and 300 K in the range of $\mu_{0}H$ = $\pm$ 7 T applied external field. Outlier points at low field are presumably due to instrumental error.}

    \label{fig:K2CuTeO4(OH)2_H2O_XvT_100Oe}
\end{figure}

\par Magnetic susceptibility measurements of \mbox{K$_{2}$CuTeO$_{4}$(OH)$_{2}$ $\cdot$ H$_{2}$O} are shown in Fig. \ref{fig:K2CuTeO4(OH)2_H2O_XvT_100Oe} and display a very broad maximum at T $\approx$ 200 K indicating low dimensional behavior with no signatures of long-range magnetic order. The strong upturn at low temperatures is likely due to small paramagnetic impurities in the sample but it cannot be ruled out as intrinsic behavior due to disorder in the system. This is supported by isothermal magnetization measurments in Fig. \ref{fig:K2CuTeO4(OH)2_H2O_XvT_100Oe} which show AFM behavior over all temperatures measured with the behavior at the T = 2 K isotherm dominated by the paramagnetic tail observed in susceptibility measurements at low temperatures. As previously discussed, the only significant magnetic interaction in this system should be the $J_{1}$ through the Cu-O-O-Cu super-superexchange pathway. In order to quantify this interaction, we have fit the magnetic susceptibility to a HTSE for a uniform AFM Heisenberg chain.\cite{Johnston2000}

\begin{equation}
    \chi_{mol} = \chi_{0} + K\chi^{*}(\dfrac{T}{J_{max}})
\end{equation}

All variables are as defined in Equation \ref{eqn:alt}. The fitting procedure shown in Fig. \ref{fig:K2CuTeO4(OH)2_H2O_XvT_100Oe} used T = 90 K as the minimum temperature but the reported fitting parameter values are displayed as the average of those for which the uniform AFM chain model was stable yielding $J_{1} = 301$ K for the uniform magnetic interaction in this system. The fittings and extracted $J_{1}$ values for the range of minimum temperatures mentioned above are shown in the SI. As expected, the uniform chain model captures the behavior observed in the magnetic susceptibility very well. 

\par It is well known that in 1D chain compounds interchain interactions ($J'$), even if very weak, typically cause transitions into long-range magnetic order. The interchain interaction strength can be estimated if the ordering temperature is known. As such, we have measured the magnetic susceptibility of \mbox{K$_{2}$CuTeO$_{4}$(OH)$_{2}$ $\cdot$ H$_{2}$O} down to T = 0.4 K (shown in the SI) and no magnetic transitions are observed. For \mbox{K$_{2}$CuTeO$_{4}$(OH)$_{2}$ $\cdot$ H$_{2}$O}, $k_{B}T_{N}/J$ $<$ 1.3 $\times$ 10$^{-3}$ which is lower than the prototypical S$= \frac{1}{2}$ 1D magnet Sr$_{2}$CuO$_{3}$ value of 2.2 $\times$ 10$^{-3}$.\cite{Motoyama1996} \mbox{K$_{2}$CuTeO$_{4}$(OH)$_{2}$ $\cdot$ H$_{2}$O} is closely related to Sr$_{2}$Cu(PO$_{4}$)$_{2}$ with similar structural and magnetic motifs of isolated CuO$_{4}$ plaquettes enabling very uniform 1D behavior. Sr$_{2}$Cu(PO$_{4}$)$_{2}$ has been touted as the best realization to date of a true 1D nearest-neighbor Heisenberg system; in this system, long range order develops at 85 mK yielding $k_{B}T_{N}/J$ = 6 $\times$ 10$^{-4}$ \cite{Johannes2006} only slightly lower than what is known for \mbox{K$_{2}$CuTeO$_{4}$(OH)$_{2}$ $\cdot$ H$_{2}$O} so far. Due to its model behavior, there has been significant experimental and theoretical interest in Sr$_{2}$Cu(PO$_{4}$)$_{2}$ as a candidate to study effects beyond the Heisenberg Hamiltonian and the corresponding Bethe-ansatz\cite{Bethe1931}. \mbox{K$_{2}$CuTeO$_{4}$(OH)$_{2}$ $\cdot$ H$_{2}$O} is positioned to be a candidate for additional study along similar lines. 

\par Additional characterization is needed to determine the ordering temperature in \mbox{K$_{2}$CuTeO$_{4}$(OH)$_{2}$ $\cdot$ H$_{2}$O} and thereby estimate the magnitude of $J'$ intrachain interactions. Neutron diffraction measurements would also be valuable. With the expectation of a $k_{B}T_{N}/J$ ratio very close to or smaller than in Sr$_{2}$Cu(PO$_{4}$)$_{2}$, an in-depth theoretical investigation via applying exact diagonalization of the Bethe-ansatz to fit the magnetic susceptibility would allow for an investigation of effects beyond the Heisenberg Hamiltonian (such as disorder, vacancies, broken chains, etc). In total, \mbox{K$_{2}$CuTeO$_{4}$(OH)$_{2}$ $\cdot$ H$_{2}$O} is distinct in the context of the K-Cu-Te-O(H) phases reported in the explored hydroflux synthesis regime. The reduced q(Cu) enables the synthesis of a more structurally and magnetically 1D motif than observed with increased q(Cu). This 1D motif makes \mbox{K$_{2}$CuTeO$_{4}$(OH)$_{2}$ $\cdot$ H$_{2}$O} a candidate for the most ideal model nearest-neighbor only Heisenberg AFM system to date.

\section{Conclusions}

In conclusion, we have synthesized three new and one previously reported layered phase in the K-Cu-Te-O(H) phase space using a hydroflux method. The three phases formed from a hydroflux melt with q(Cu) = 1, \mbox{K$_{2}$Cu$_{2}$TeO$_{6}$ $\cdot$ H$_{2}$O},  K$_{2}$Cu$_{2}$TeO$_{6}$, and \mbox{K$_{6}$Cu$_{9}$Te$_{4}$O$_{24}$ $\cdot$ 2 H$_{2}$O}, all contain Cu$^{2+}$ honeycomb layers formed from CuO$_{4}$ plaquettes tiled with TeO$_{6}$ octahedra, with interlayer K$^{+}$ ions and increasing incorporation of interlayer water as q(K) is increased. Each phase shows varying degrees of CuO$_{6}$ octahedral distortion; in all cases, an apical distortion leads to an effective square planar CuO$_{4}$ plaquette. Distinct structural distortions in each phase lead to distinct tilt sequences within the honeycomb lattice which limit some magnetic interactions and enhance others. \mbox{K$_{2}$Cu$_{2}$TeO$_{6}$ $\cdot$ H$_{2}$O}, containing edge-sharing CuO$_{4}$ plaquettes, does not order down to T = 2 K and displays alternating chain Heisenberg AFM behavior. More detailed characterization using neutron spectroscopy needs to be done to characterize the magnetic interactions and understand its magnetic behavior in the context of other honeycomb and alternating chain Heisenberg systems. K$_{2}$Cu$_{2}$TeO$_{6}$, which is structurally similar to \mbox{K$_{2}$Cu$_{2}$TeO$_{6}$ $\cdot$ H$_{2}$O} but contains only corner-sharing CuO$_{4}$ plaquettes, displays AFM order at T$_{N}$ = 100 K which likely results from the two-dimensional nature of the magnetic interactions driven by the structural distortions. The comparison of these two phases provides a model test case to understand how an additional dimension of order can be induced (in this case, from 1D to 2D interactions). Polar phase \mbox{K$_{6}$Cu$_{9}$Te$_{4}$O$_{24}$ $\cdot$ 2 H$_{2}$O}, which contains both edge- and corner-sharing CuO$_{4}$ plaquettes within the honeycomb layer, as well as a CuO$_{4}$ plaquette which bridge the layers, shows AFM order at T$_{N}$ = 6.5 K and is generally described well by a nearest-neighbor honeycomb model. The fourth phase investigated in this work, \mbox{K$_{2}$CuTeO$_{4}$(OH)$_{2}$ $\cdot$ H$_{2}$O}, formed with q(K)= 2, 5 and q(Cu) = 0.25 and contains structurally and magnetically isolated chains of CuO$_{4}$ plaquettes leading to uniform chain Heisenberg AFM behavior with no magnetic order down to T = 0.4 K. This phase is a candidate model nearest-neighbor only Heisenberg system. The various CuO$_{4}$ plaquette tilting patterns in these phases lead to complex magnetic interactions which require further study. Since all of the new phases formed a honeycomb motif, which is a 1/3 depletion of the triangular lattice, further work could examine tuning q(Cu) between 1 and 4, in an attempt to tune the depletion to a smaller ratio, possibly stabilizing other important magnetic motifs such as the kagome (1/4 depletion) and maple leaf (1/7 depletion) lattices. Additional tuning of the hydroflux within this phase space and related phase spaces could lead to the discovery of new structural and magnetic motifs, and further investigation would aid in the specific understanding of the mechanisms of hydroflux synthesis.

\section{Acknowledgments}

\par The MPMS3 system used for magnetic characterization was funded by the National Science Foundation, Division of Materials Research, Major Research Instrumentation Program, under Grant \#1828490. DFT calculations were performed using computational resources of the Maryland Advanced Research Computing Center and the Advanced Research Computing at Hopkins (ARCH) Rockfish cluster. E.Z. acknowledges support from the Sweeney Family Postdoctoral Fellowship.

\bibliography{main.bib}

\end{document}

\begin{center}

\textbf{Supplementary Material for\\ ``Hydroflux-Controlled Growth of Magnetic K-Cu-Te-O(H) Phases''}

\end{center}

\begin{longtable}{lcccccc}
\caption[\pagewidth]{Structural, lattice, and anisotropic displacement parameters for all K-Cu-Te-O(H) compounds discussed in this work determined at T = 213(2) K. Errors for all crystallographic sites and isotropic displacement parameters are shown where applicable. Occupancy of all sites were fixed to unity.} 
%\begin{tabular}{lcccccc} 						
 \hline Compound & Atom & Wyckoff Site	& $x$ &	$y$ & $z$ & $U_{eq}$({\AA$^{2}$})\\ \hline	
 K$_{2}$Cu$_{2}$TeO$_{6}$ & Te1 & 2d & 0.5000  & 0.5000 & 0.0000 & 0.00464(4)\\
 & Cu1 & 4e & 0.55324(3) & 0.33883(2) & 0.49408(3) & 0.00549(4)\\
 & K1  & 4e & 0.95692(5)	& 0.63878(4) & 0.71935(7) & 0.00159(7)\\
 & O1  & 4e & 0.70330(16)& 0.38680(11) & -0.1389(2) & 0.00834(18)\\
 & O2  & 4e & 0.38855(16)& 0.32002(11) & 0.1066(2) & 0.00678(17) \\
 & O3  & 4e & 0.69316(16)& 0.49566(11) & 0.3443(2) & 0.00710(18) \vspace{0.25in} \\

 K$_{2}$Cu$_{2}$TeO$_{6}$ $\cdot$ H$_{2}$O & Te1 & 4a & 0.0000  & 0.0000 & 0.0000 & 0.00541(9)\\
 & Cu1 & 8e & -0.16592(4)& 0.5000      & 0.0000	& 0.00795(11)\\
 & K1  & 8g & 0.32331(9) & -0.27118(14)& 0.2500	& 0.001432(15)\\
 & O1  & 8f & 0.0000     & 0.3111(4)	  & 0.6377(18) & 0.0079(4)\\
 & O2  & 16h& 0.16095(18)& -0.991(3)   & 0.9193(14) & 0.0094(3) \\
 & O3  & 4c & 0.5000	    & 0.1180(7)	  & 0.2500    & 0.00178(8) \\
 & H3  & 8f & 0.5000    & 0.222(7)    & 0.294(3)  & 0.027  \\\vspace{0.25in} \\

 K$_{6}$Cu$_{9}$Te$_{4}$O$_{24}$ $\cdot$ 2 H$_{2}$O & Te1 & 4b & 0.22136(3) & 0.88584(3)	& 0.73568(4)	& 0.00565(8) \\
& Te2 & 4b & 0.26264(3) & 0.61318(3)	& 0.26180(3)	& 0.00584(8) \\
& Cu1	& 4b & 0.25680(10) & 0.87150(8) & 0.41691(9)	& 0.00675(17) \\
& Cu2	& 4b & 0.25777(10) & 0.37187(8)	& 0.42068(9)	& 0.00754(18) \\
& Cu3	& 4b & 0.24752(10) & 0.86906(8)	& 1.08244(10)	& 0.00718(16) \\
& Cu4	& 4b & 0.23121(8) & 0.62091(8)	& 0.58526(9)	& 0.00676(19) \\
& Cu5	& 2a & 0.0000     & 0.85093(9)	& 0.82003(12)	& 0.0096(2) \\
& K1	& 2a & 0.5000     & 0.51091(18) & -0.0954(2)	& 0.0150(4) \\
& K2	& 2a & 0.0000     & 1.04277(19)	& 1.0956(2)	    & 0.0173(4) \\
& K3	& 2a & 0.0000     & 0.8348(2)	& 0.4257(3)	    & 0.0249(5) \\
& K4	& 2a & 0.0000	  & 0.55922(16)	& 0.7692(2)	    & 0.0166(4) \\
& K5	& 2a & 0.5000	  & 0.86856(18)	& 0.1610(2)	    & 0.0166(4) \\
& K6	& 2a & 0.5000	  & 0.6424(2)	    & 0.5593(3)	    & 0.0256(5) \\
& O1	& 4b & 0.2959(4) & 0.8047(4)	& 0.8951(5)	& 0.0085(9) \\
& O2	& 4b & 0.3116(3) & 1.0359(3)	& 0.7509(5)	& 0.0068(7) \\
& O3	& 4b & 0.1711(3) & 0.9729(4)	& 0.5604(5)	& 0.0080(8) \\
& O4	& 4b & 0.3117(3) & 0.5513(3)	& 0.7488(5)	& 0.0093(7) \\
& O5	& 4b & 0.1208(3) & 0.9665(4)	& 0.8621(4)	& 0.0095(8) \\
& O6	& 4b & 0.1615(4) & 0.6711(4)	& 0.4017(5)	& 0.0091(8) \\
& O7	& 4b & 0.1153(3) & 0.7577(3)	& 0.7235(5)	& 0.0105(8) \\
& O8	& 4b & 0.1838(4) & 0.6970(4)	& 0.1050(5)	& 0.0081(8) \\
& O9	& 4b & 0.3089(4) & 0.7952(4)	& 0.5991(5)	& 0.0087(8) \\
& O10	& 4b & 0.3413(3) & 0.7641(3)	& 0.2843(4)	& 0.0094(8) \\
& O11	& 4b & 0.3496(4) & 0.5442(4)	& 0.1100(5)	& 0.0107(8) \\
& O12	& 4b & 0.3394(4) & 0.5286(4)	& 0.4147(5)	& 0.0101(9) \\
& O13	& 2a & 0.0000    & 0.7873(7)	& 1.0594(9)	& 0.0267(17) \\
& O14	& 2a & 0.5000    & 0.7636(6)	& 0.8153(7)	& 0.0202(14) \\
& H13   & 4b & 0.050(5)  & 0.750(6)     & 1.091(9)  & 0.044 \\
& H14   & 4b & 0.452(5)  & 0.769(6)     & 0.869(7)  & 0.032 \\
\vspace{0.25in} \\

 K$_{2}$CuTeO$_{4}$(OH)$_{2}$ $\cdot$ H$_{2}$O & 
   Te1 & 4a & 0.24996(2) & 0.25356(5) & 0.49998(2) & 0.00793(5\\
 & Cu1 & 4a & 0.50128(12) & 0.49968(10) &  0.50008(11) & 0.01041(7)\\
 & K1  & 4a & 0.17355(15) & 0.2766(2) & 0.18910(11) & 0.0201(3)\\
 & K2  & 4a & 0.29089(15) & 0.23583(17) &  0.81771(11) & 0.0183(3)\\
 & O1  & 4a & 0.3773(5) & 0.3619(6) & 0.3920(4) & 0.0114(6) \\
 & O2  & 4a & 0.1250(4) & 0.1411(6) & 0.6088(4) & 0.0103(6) \\
 & O3  & 4a & 0.3696(4) & 0.3859(6) & 0.6075(4) & 0.0106(6) \\
 & O4  & 4a & 0.1327(5) & 0.1173(6) & 0.3935(4) & 0.0120(7) \\
 & O5  & 4a & 0.1472(4) & 0.5257(5) & 0.4866(4) & 0.0140(7) \\
 & H5  & 4a & 0.109(6) & 0.522(9) & 0.429(4) & 0.021 \\
 & O6  & 4a & 0.3603(4) & -0.0207(5) & 0.5102(4) & 0.0146(7) \\
 & H6  & 4a & 0.445(4) & -0.016(9) & 0.501(5) & 0.022 \\
 & O7  & 4a & 0.4719(4) & 0.1027(5) & 0.2328(3) & 0.0210(7) \\
 & H7  & 4a & 0.448(6) & 0.179(10) & 0.286(4) & 0.032 \\
 & H8  & 4a & 0.516(6) & 0.181(10) & 0.192(4) & 0.032 \\
 
\hline																										
%\end{tabular}
\label{scxrd_params}
\end{longtable}

\begin{figure*}[h!]
    \centering
    \includegraphics[scale = 0.75]{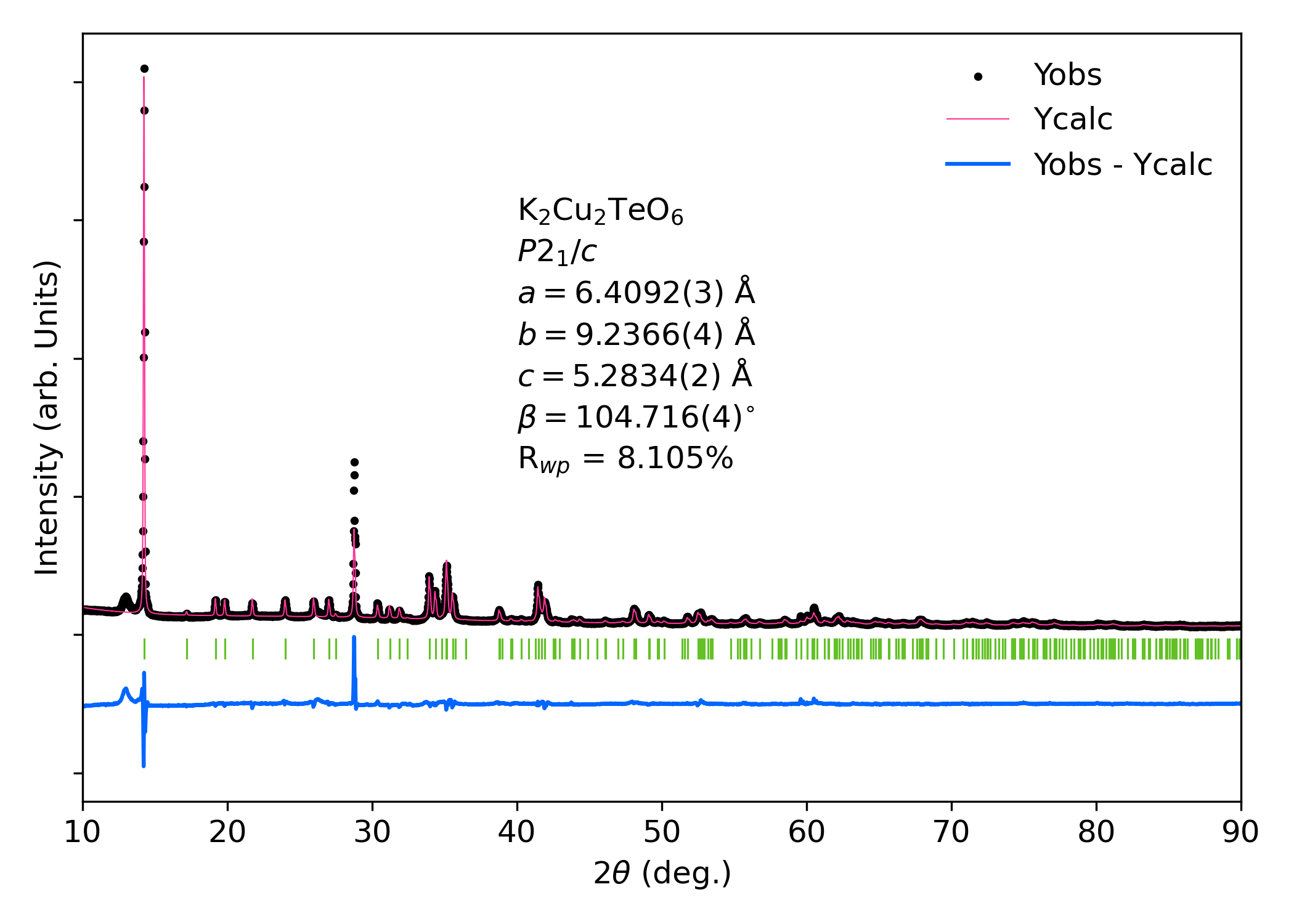}
    \caption{Powder X-ray diffraction of phase pure K$_{2}$Cu$_{2}$TeO$_{6}$ and Rietveld refinement using the solved structure from single crystal X-ray diffraction. Extracted lattice parameters are shown and are in close agreement to single crystal measurements. The broad peak at $2\theta \approx 12^{\circ}$ is likely due to hydration. Despite this hydration, note that the formation of the distinct hydrated phase K$_{2}$Cu$_{2}$TeO$_{6}$ $\cdot$ H$_{2}$O is not observed even after washing with water to remove excess KOH.} 
\end{figure*}

\begin{figure*}[h!]
    \centering
    \includegraphics[scale = 0.75]{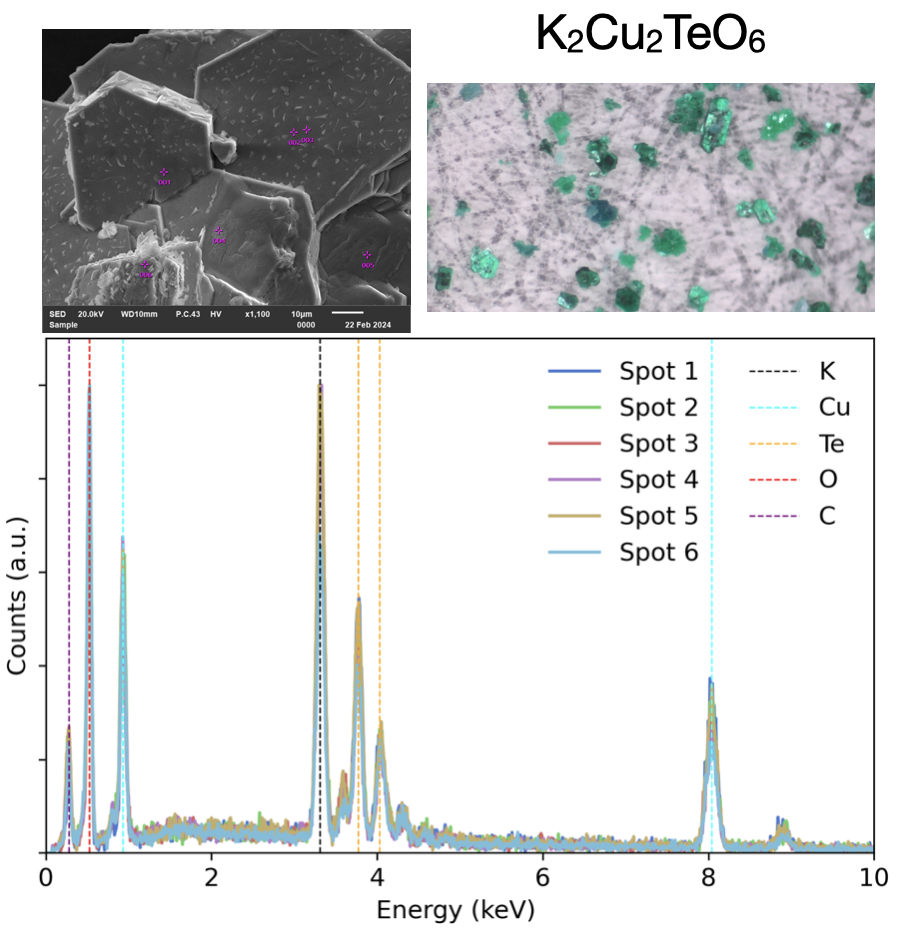}
    \caption{SEM images, microscope images, and EDS spectra of as-recovered K$_{2}$Cu$_{2}$TeO$_{6}$. Carbon present in the EDS spectrum likely arises from the carbon tape used to mount the samples.} 
\end{figure*}

\begin{table}[h!]
\begin{tabular}{|c|c|c|c|c|c|c|c|c|}
\hline
K$_{2}$Cu$_{2}$TeO$_{6}$ & K wt\% & Cu wt\% & Te wt\% & O wt\% & K at\% & Cu at\% & Te at\% & O at\% \\ \hline
Spot 1                   & 16.83  & 29.60   & 33.22   & 20.35  &        &         &         &        \\ \hline
Spot 2                   & 17.73  & 26.98   & 30.26   & 25.02  &        &         &         &        \\ \hline
Spot 3                   & 19.66  & 24.55   & 30.54   & 25.25  &        &         &         &        \\ \hline
Spot 4                   & 18.74  & 23.63   & 30.66   & 26.97  &        &         &         &        \\ \hline
Spot 5                   & 18.24  & 27.22   & 33.04   & 21.50  &        &         &         &        \\ \hline
Spot 6                   & 14.50  & 25.11   & 31.25   & 24.70  &        &         &         &        \\ \hline
Average                  & 17(2)  & 25(3)   & 31(1)   & 26(2)  & 14(2)  & 13(3)   & 7(2)    & 65(1)  \\ \hline
\end{tabular}
\caption{Quantitative EDS spectra data determining the K, Cu, Te, and O composition at different points of the K$_{2}$Cu$_{2}$TeO$_{6}$ measured sample. The atomic percent of K, Cu, Te, and O is calculated from the reported average weight percent. The calculated ratio of K:Cu:Te is close to 2:2:1 in agreement with the compound's stoichiometry. Increased oxygen content may be due to hydration.}
\end{table}

\begin{figure*}[h!]
    \centering
    \includegraphics[scale = 0.75]{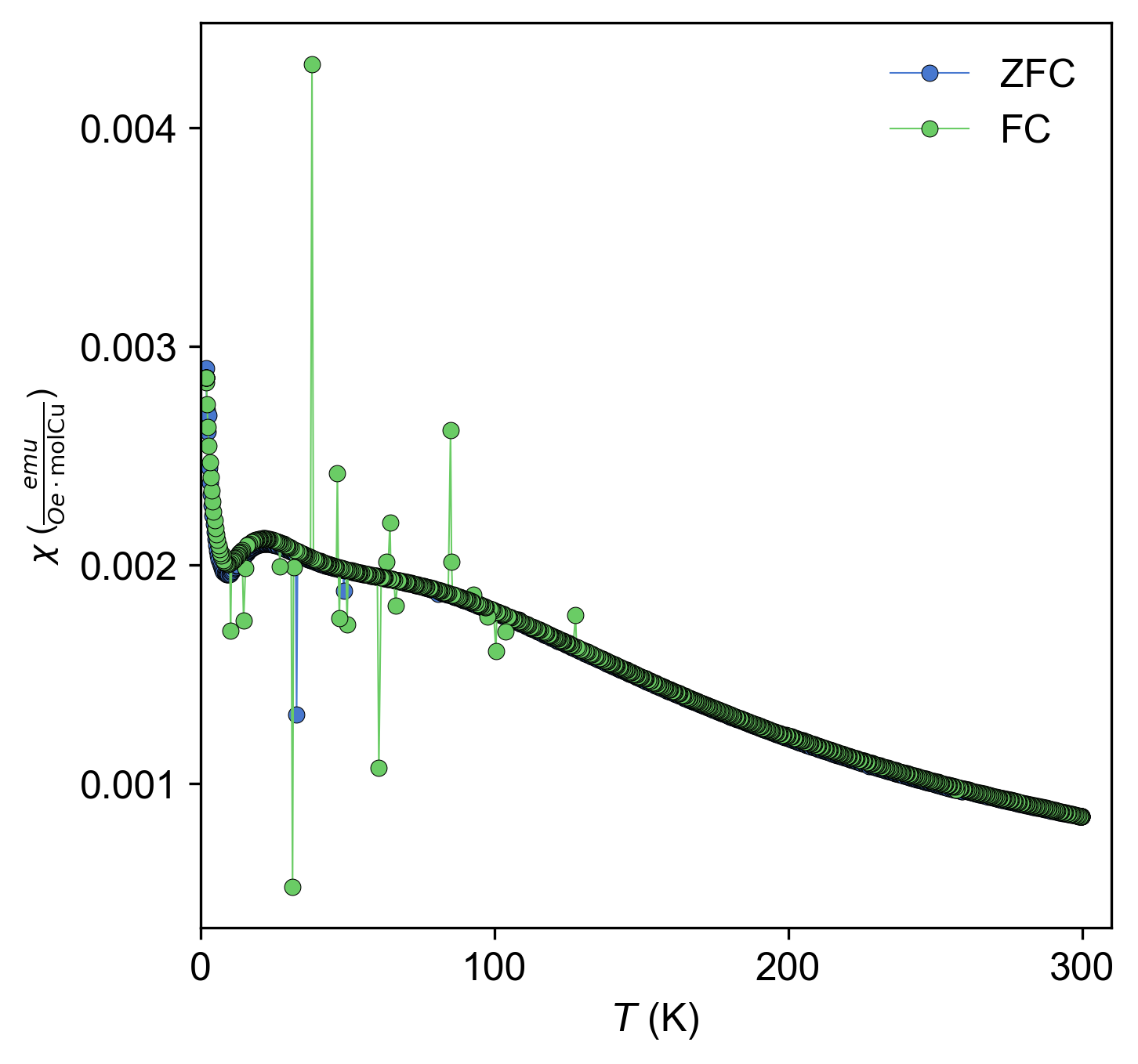}
    \caption{Molar magnetic susceptibility of K$_{2}$Cu$_{2}$TeO$_{6}$ measured in ZFC and FC regimes with an external field of $\mu_{0}H$ = 1 T showing low-dimensional behavior and a suppression of the antiferromagentic transition observed at T = $\approx$ 100 K at H = 100 Oe external field. } 
\end{figure*}

\begin{figure*}[h!]
    \centering
    \includegraphics[scale = 0.75]{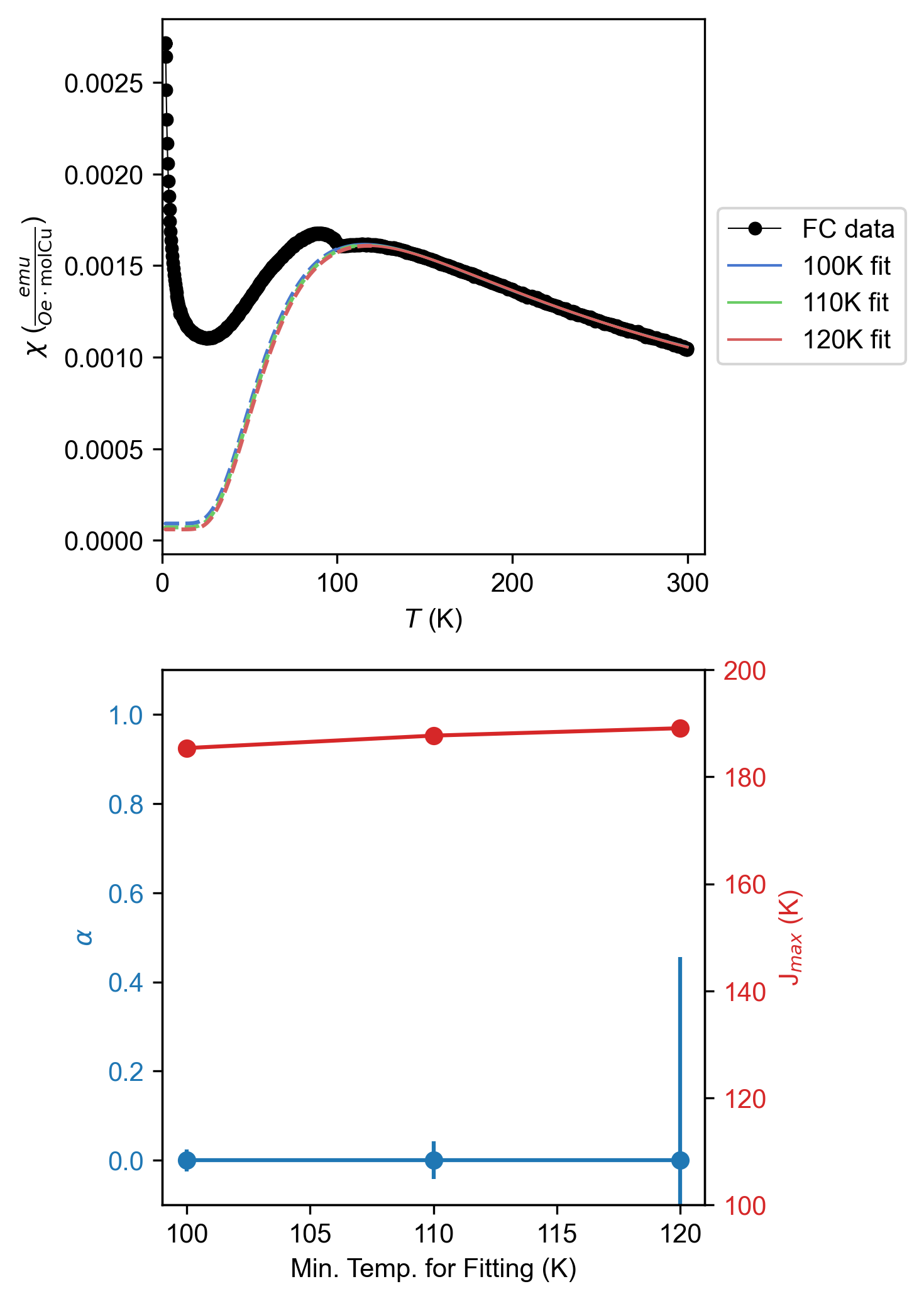}
    \caption{(top) Alternating chain AFM Heisenberg fitting of the magnetic susceptibility of K$_{2}$Cu$_{2}$TeO$_{6}$ choosing different minimum temperatures for the high-temperature range of the fitting procedure. (bottom) Resulting fit parameters, $\alpha$ and $J_{max}$, from the choice of different minimum temperature cut-offs in the fitting procedure. Only temperature ranges for which the fit parameters remained stable were chosen to be included. } 
\end{figure*}

\begin{figure*}[h!]
    \centering
    \includegraphics[scale = 0.75]{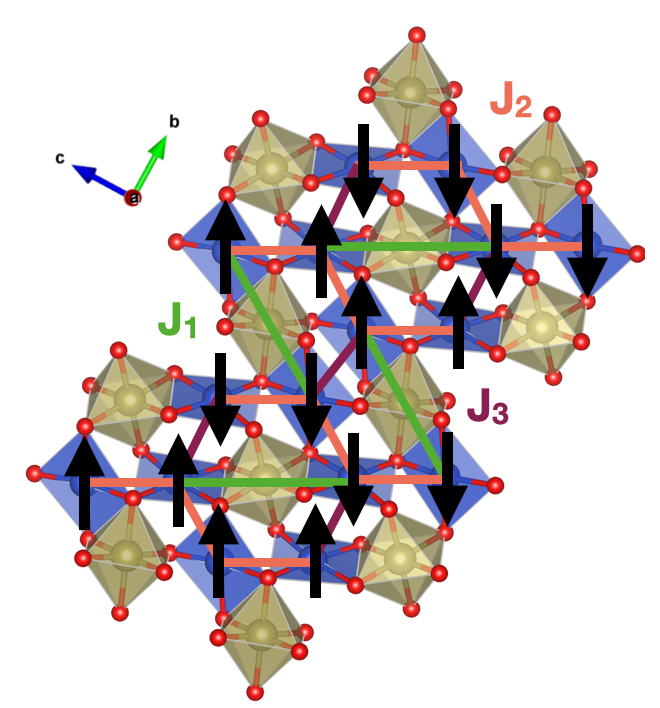}
    \caption{Proposed k = (0, 0, 0) AFM structure of K$_{2}$Cu$_{2}$TeO$_{6}$ adhering to the magnitude and sign of $J_1$, $J_2$, and $J_3$ of the model discussed in the main text.} 
\end{figure*}

\begin{figure*}[h!]
    \centering
    \includegraphics[scale = 0.75]{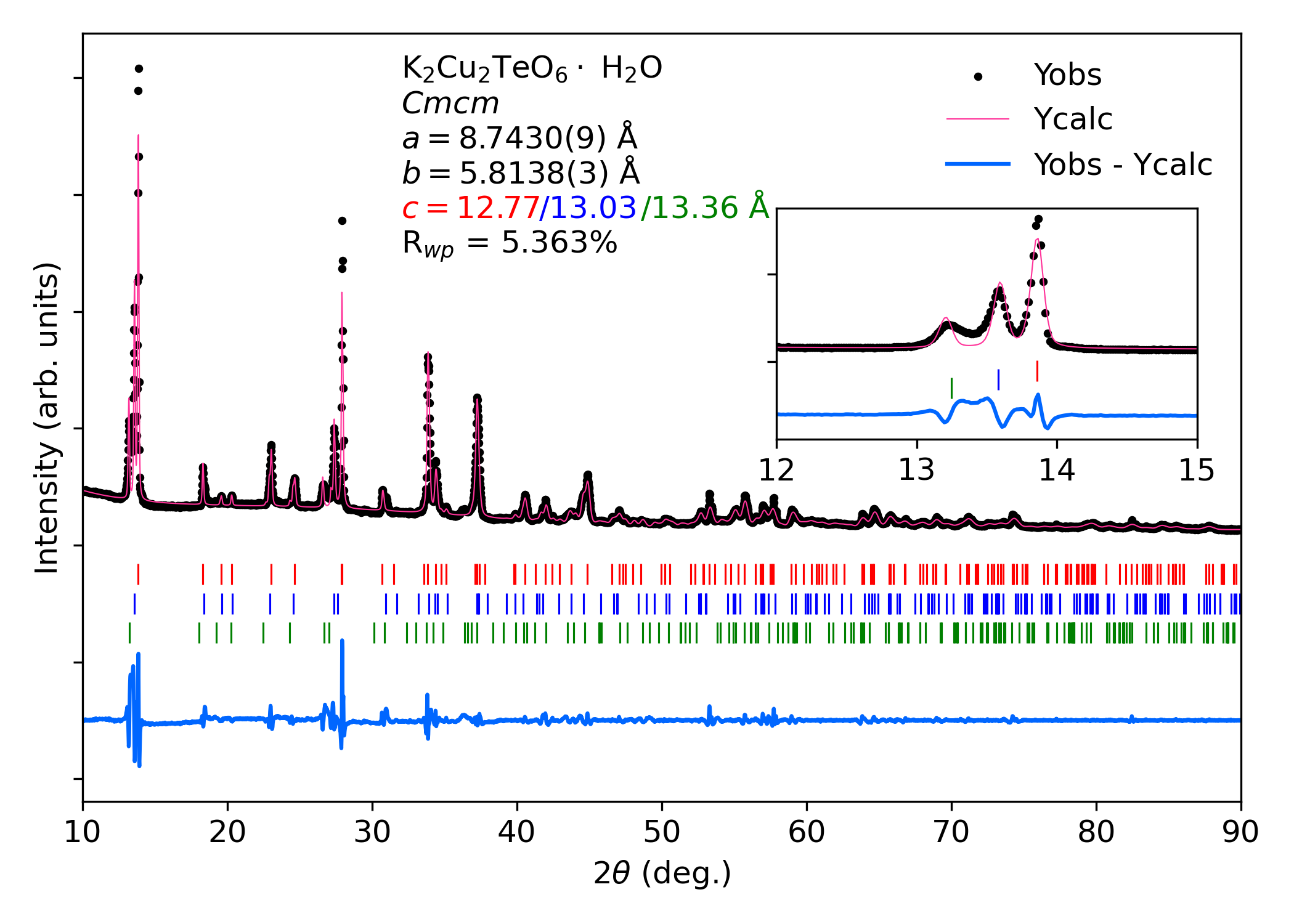}
    \caption{Powder X-ray diffraction of K$_{2}$Cu$_{2}$TeO$_{6}$ $\cdot$ H$_{2}$O and Pawley refinement using the solved structure from single crystal X-ray diffraction. Pawley refinement was chosen due to the appearance of multiple phases with similar lattice parameters but altered $c$-axis lattice parameters indicating slightly different interlayer spacing likely due to effects of hydration over time. Extracted lattice parameters are shown and are in close agreement to single crystal measurements.} 
\end{figure*}

\begin{figure*}[h!]
    \centering
    \includegraphics[scale = 0.5]{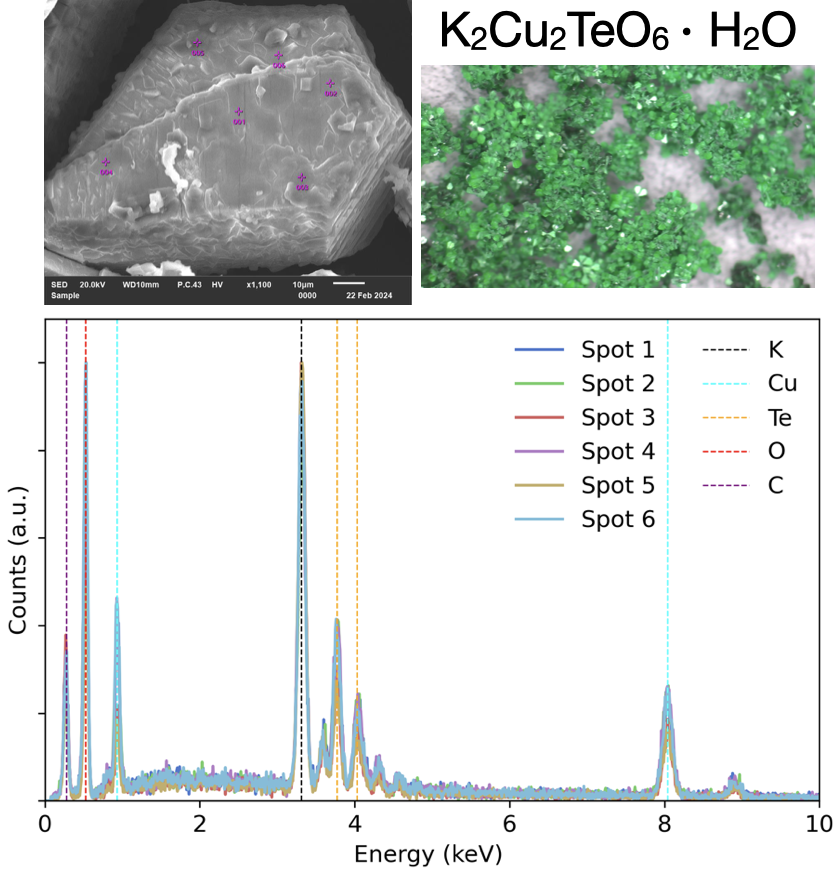}
    \caption{SEM images, microscope images, and EDS spectra of as-recovered K$_{2}$Cu$_{2}$TeO$_{6}$ $\cdot$ H$_{2}$O. Carbon present in the EDS spectrum likely arises from the carbon tape used to mount the samples.} 
\end{figure*}

\begin{table}[h!]
\begin{tabular}{|c|c|c|c|c|c|c|c|c|}
\hline
K$_{2}$Cu$_{2}$TeO$_{6} \cdot$ H$_{2}$O & K wt\% & Cu wt\% & Te wt\% & O wt\% & K at\% & Cu at\% & Te at\% & O at\% \\ \hline
Spot 1                                  & 19.45  & 22.34   & 29.29   & 28.92  &        &         &         &        \\ \hline
Spot 2                                  & 17.43  & 24.25   & 30.28   & 28.04  &        &         &         &        \\ \hline
Spot 3                                  & 24.85  & 19.36   & 24.36   & 31.43  &        &         &         &        \\ \hline
Spot 4                                  & 17.90  & 24.92   & 30.01   & 27.17  &        &         &         &        \\ \hline
Spot 5                                  & 27.34  & 20.53   & 25.86   & 26.27  &        &         &         &        \\ \hline
Spot 6                                  & 18.45  & 23.40   & 28.58   & 29.57  &        &         &         &        \\ \hline
Average                                 & 21(4)  & 22(2)   & 28(2)   & 28(2)  & 18(4)  & 12(2)   & 7(2)    & 61(2)  \\ \hline
\end{tabular}
\caption{Quantitative EDS spectra data determining the K, Cu, Te, and O composition at different points of the K$_{2}$Cu$_{2}$TeO$_{6}$ $\cdot$ H$_{2}$O measured sample. The atomic percent of K, Cu, Te, and O is calculated from the reported average weight percent. The calculated ratio of K:Cu:Te is close to 2:2:1 in agreement with the compound's stoichiometry. Increased oxygen content may be due to hydration.}
\end{table}

\begin{figure*}[h!]
    \centering
    \includegraphics[scale = 0.75]{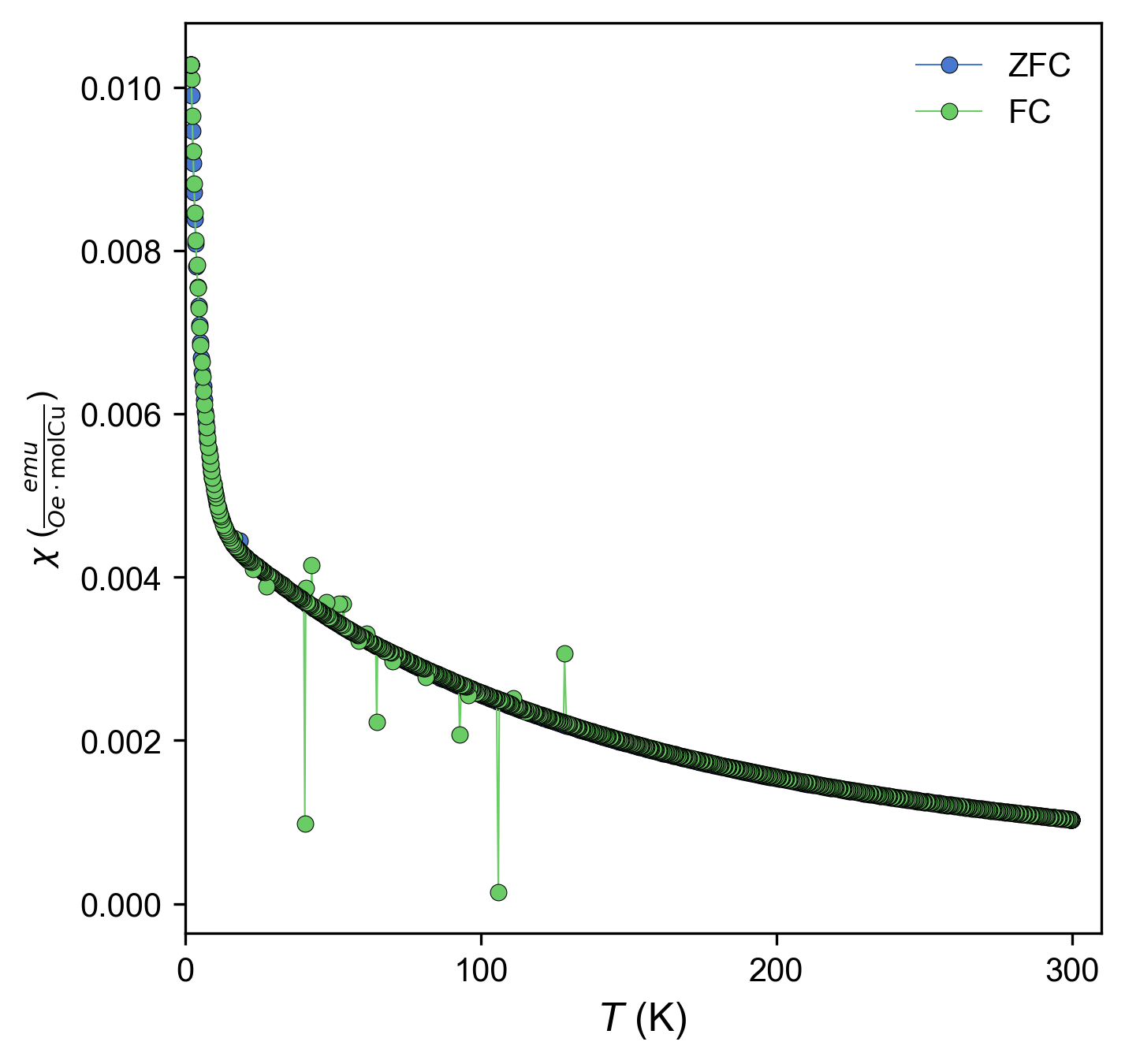}
    \caption{Molar magnetic susceptibility of K$_{2}$Cu$_{2}$TeO$_{6}$ $\cdot$ H$_{2}$O measured in ZFC and FC regimes with an external field of $\mu_{0}H$ = 1 T showing low-dimensional behavior.} 
\end{figure*}

\begin{figure*}[h!]
    \centering
    \includegraphics[scale = 0.75]{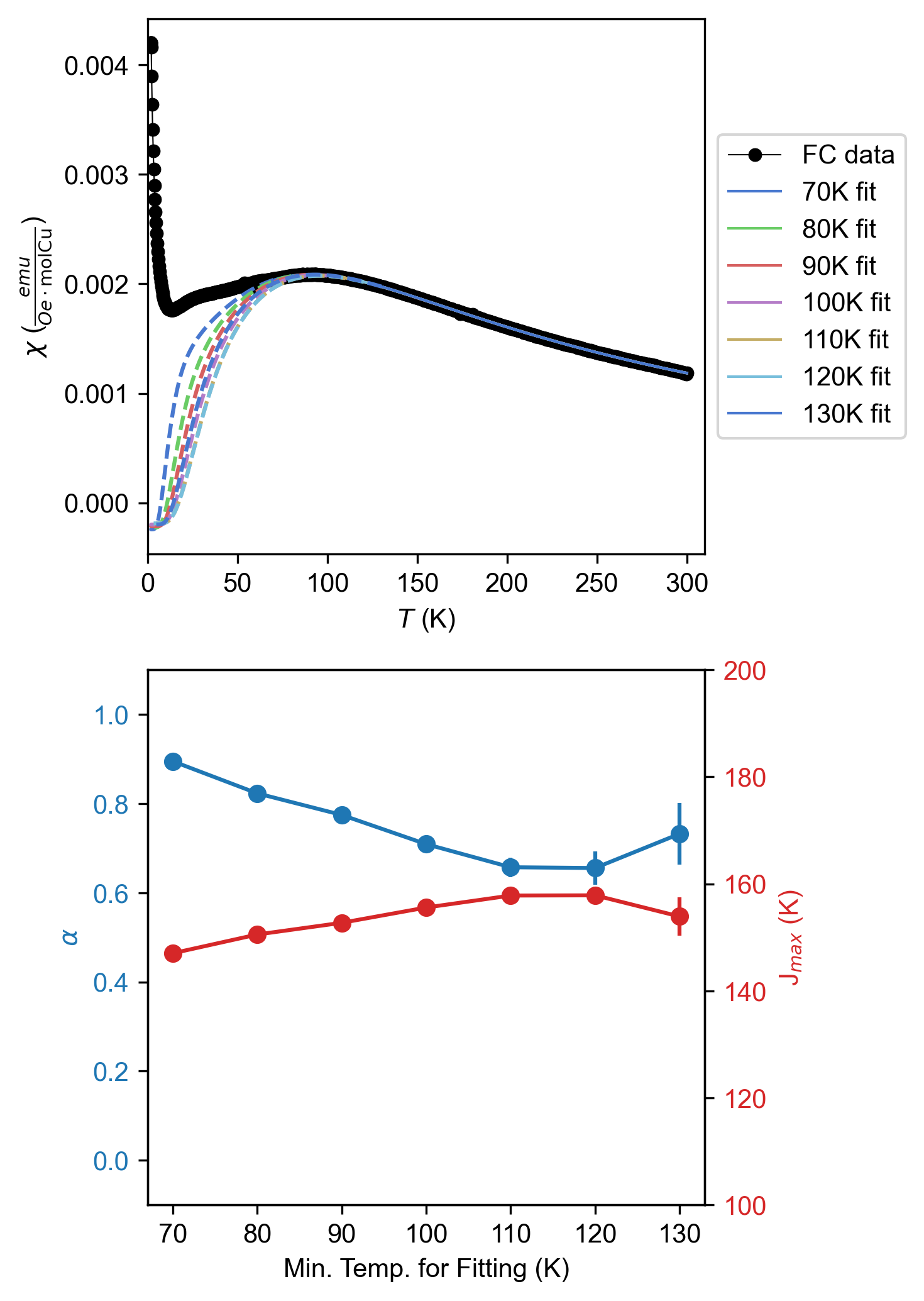}
    \caption{(top) Alternating chain AFM Heisenberg fitting of the magnetic susceptibility of K$_{2}$Cu$_{2}$TeO$_{6}$ $\cdot$ H$_{2}$O choosing different minimum temperatures for the high-temperature range of the fitting procedure. (bottom) Resulting fit parameters, $\alpha$ and $J_{max}$, from the choice of different minimum temperature cut-offs in the fitting procedure. Only temperature ranges for which the fit parameters remained stable were chosen to be included.} 
\end{figure*}

\begin{figure*}[h!]
    \centering
    \includegraphics[scale = 0.75]{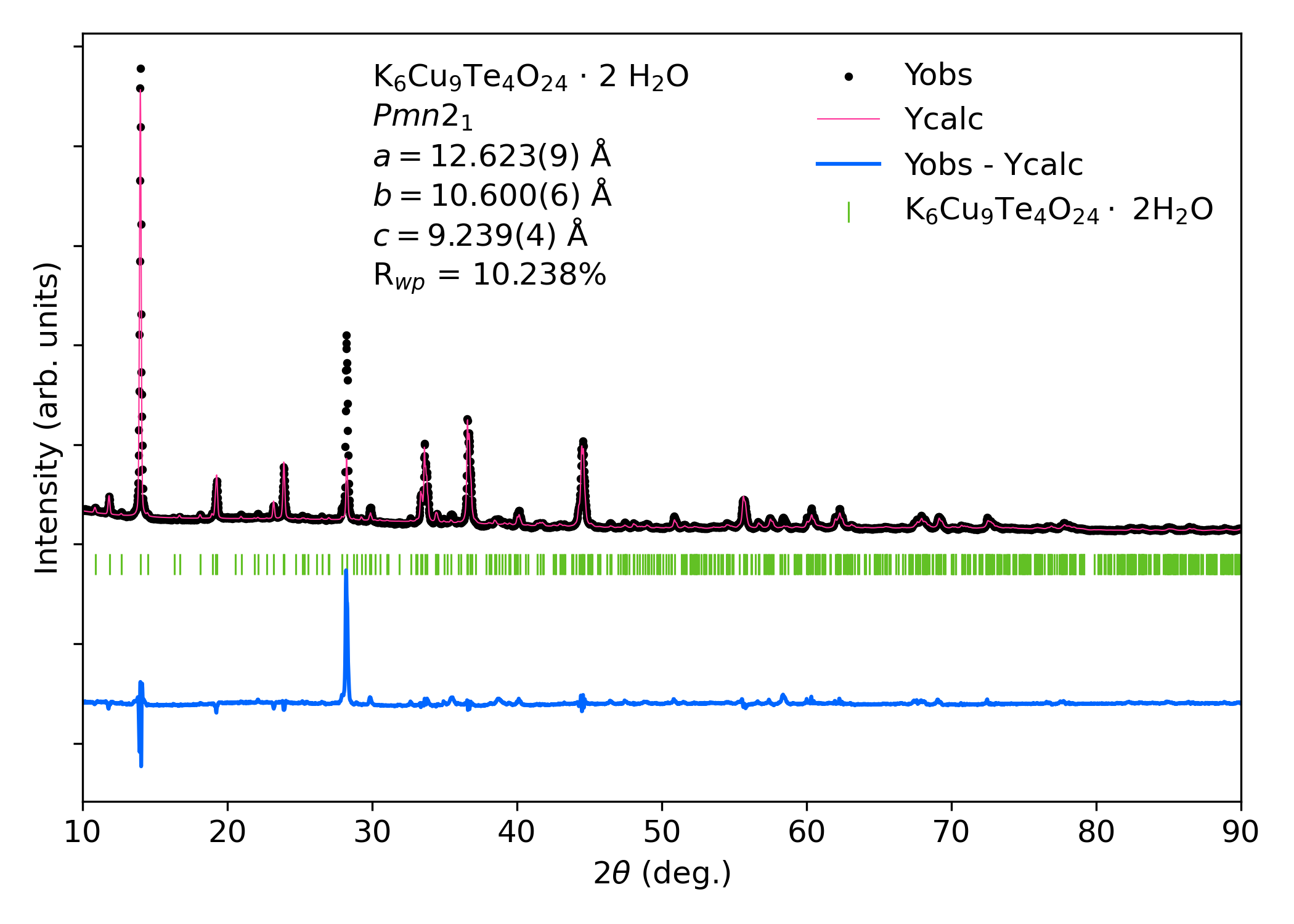}
    \caption{Powder X-ray diffraction of phase pure K$_{6}$Cu$_{9}$Te$_{4}$O$_{24}$ $\cdot$ 2 H$_{2}$O and Rietveld refinement using the solved structure from single crystal X-ray diffraction. Extracted lattice parameters are shown and are in close agreement to single crystal measurements.} 
\end{figure*}

\begin{figure*}[h!]
    \centering
    \includegraphics[scale = 0.75]{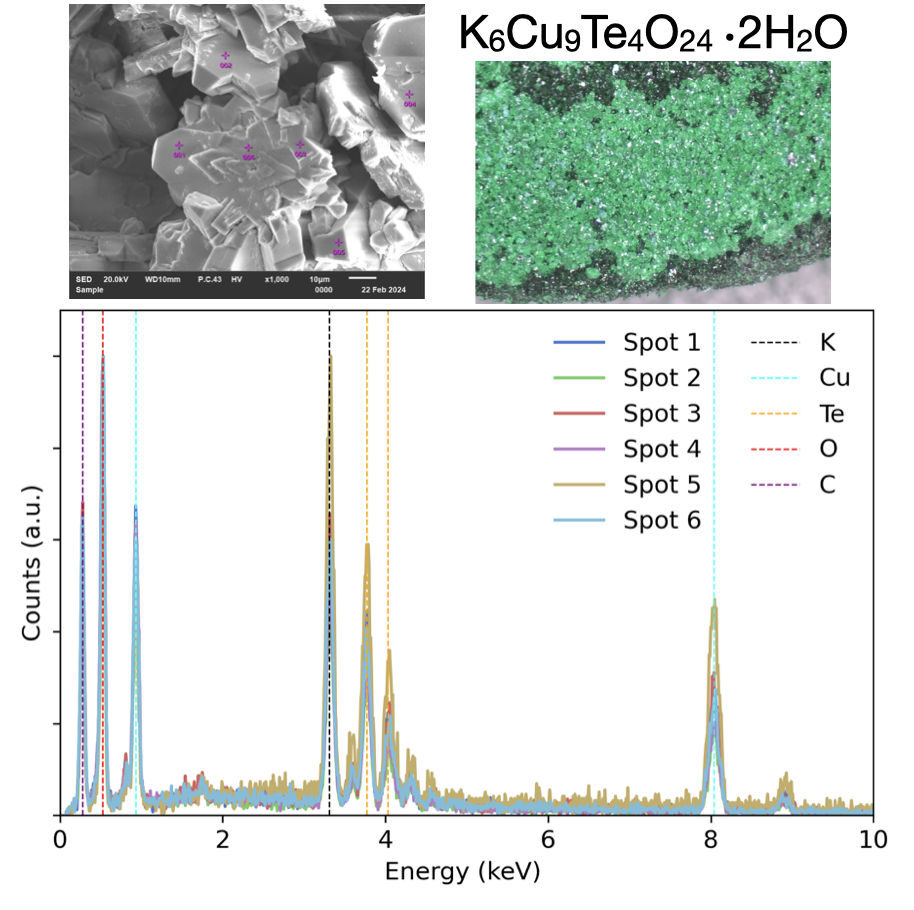}
    \caption{SEM images, microscope images, and EDS spectra of as-recovered K$_{6}$Cu$_{9}$Te$_{4}$O$_{24}$ $\cdot$ 2 H$_{2}$O. Carbon present in the EDS spectrum likely arises from the carbon tape used to mount the samples.} 
\end{figure*}

\begin{table}[h!]
\begin{tabular}{|c|c|c|c|c|c|c|c|c|}
\hline
K$_{6}$Cu$_{9}$Te$_{4}$O$_{24}$ $\cdot$ 2 H$_{2}$O & K wt\% & Cu wt\% & Te wt\% & O wt\% & K at\% & Cu at\% & Te at\% & O at\% \\ \hline
Spot 1                                             & 13.65  & 27.93   & 31.50   & 26.93  &        &         &         &        \\ \hline
Spot 2                                             & 12.23  & 25.74   & 28.99   & 33.04  &        &         &         &        \\ \hline
Spot 3                                             & 13.63  & 28.19   & 30.91   & 27.27  &        &         &         &        \\ \hline
Spot 4                                             & 13.14  & 25.63   & 29.96   & 31.27  &        &         &         &        \\ \hline
Spot 5                                             & 14.80  & 35.54   & 34.22   & 15.44  &        &         &         &        \\ \hline
Spot 6                                             & 13.37  & 28.50   & 29.69   & 28.45  &        &         &         &        \\ \hline
Average                                            & 13(1)  & 28(4)   & 31(2)   & 27(6)  & 12(1)  & 16(4)   & 10(2)   & 61(6)  \\ \hline
\end{tabular}
\caption{Quantitative EDS spectra data determining the K, Cu, Te, and O composition at different points of the K$_{6}$Cu$_{9}$Te$_{4}$O$_{24}$ $\cdot$ 2 H$_{2}$O measured sample. The atomic percent of K, Cu, Te, and O is calculated from the reported average weight percent. The calculated ratio of K:Cu:Te:O is close to 6:9:4:26 in agreement with the compound's stoichiometry.}
\end{table}

\begin{figure*}[h!]
    \centering
    \includegraphics[scale = 0.75]{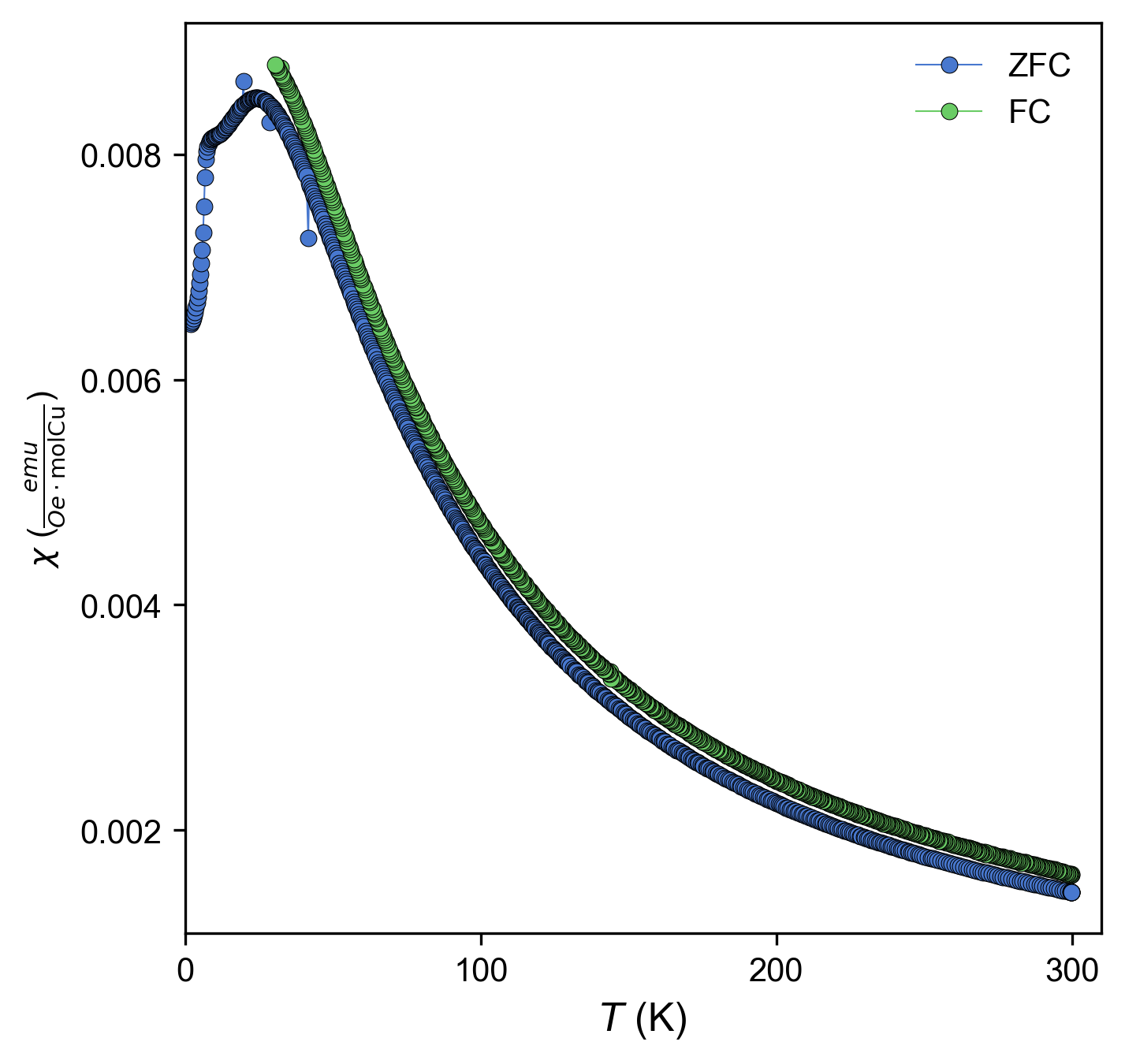}
    \caption{Molar magnetic susceptibility of K$_{6}$Cu$_{9}$Te$_{4}$O$_{24}$ $\cdot$ 2 H$_{2}$O  measured in ZFC and FC regimes with an external field of $\mu_{0}H$ = 1 T. } 
\end{figure*}

\begin{figure*}[h!]
    \centering
    \includegraphics[scale = 0.75]{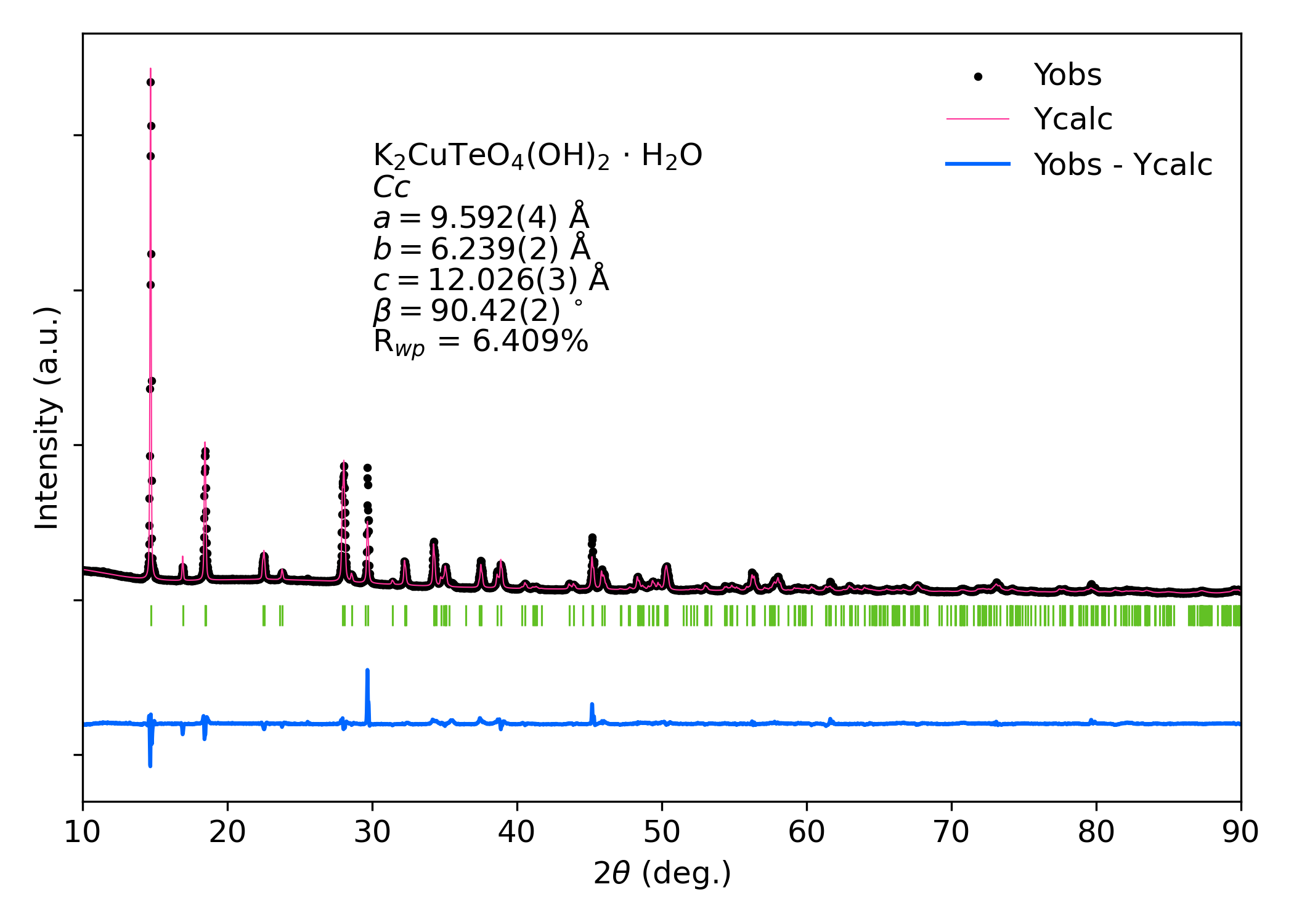}
    \caption{Powder X-ray diffraction of phase pure K$_{2}$CuTeO$_{4}$(OH)$_{2}$ $\cdot$ H$_{2}$O and Rietveld refinement using the solved structure from single crystal X-ray diffraction. Extracted lattice parameters are shown and are in close agreement to single crystal measurements.} 
\end{figure*}

\begin{figure*}[h!]
    \centering
    \includegraphics[scale = 0.75]{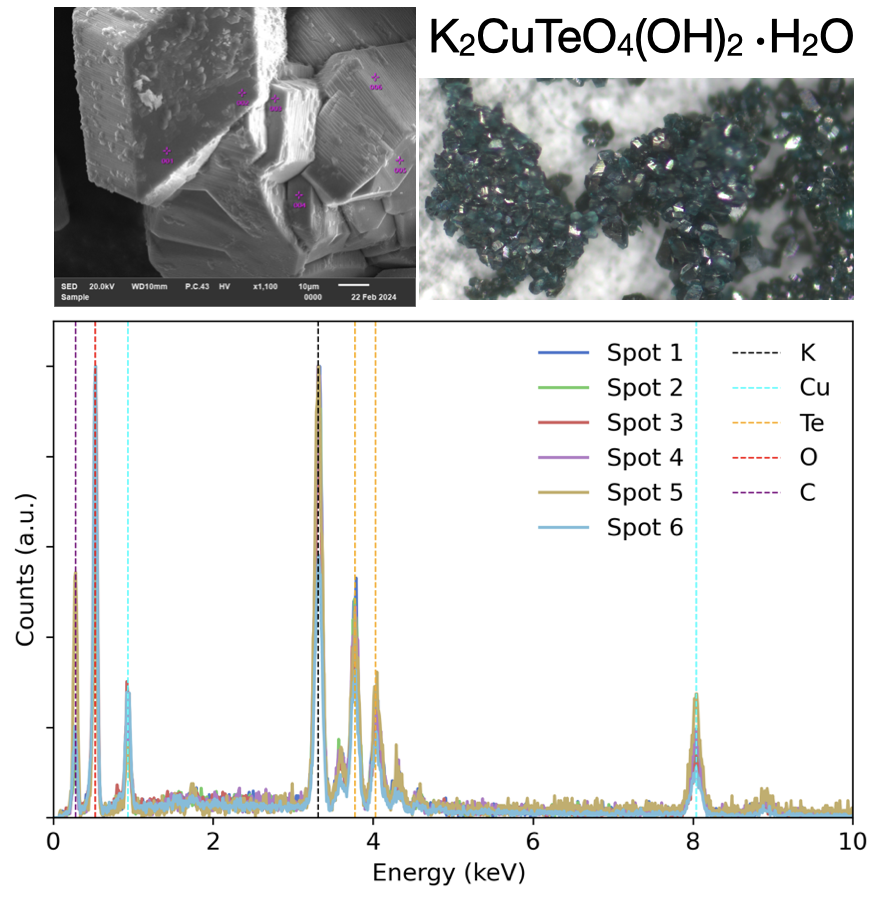}
    \caption{SEM images, microscope images, and EDS spectra of as-recovered K$_{2}$CuTeO$_{4}$(OH)$_{2}$ $\cdot$ H$_{2}$O. Carbon present in the EDS spectrum likely arises from the carbon tape used to mount the samples.} 
\end{figure*}

\begin{table}[h!]
\begin{tabular}{|c|c|c|c|c|c|c|c|c|}
\hline
K$_{2}$CuTeO$_{4}$(OH)$_{2}$ $\cdot$ H$_{2}$O & K wt\% & Cu wt\% & Te wt\% & O wt\% & K at\% & Cu at\% & Te at\% & O at\% \\ \hline
Spot 1                                          & 21.11  & 15.76   & 38.43   & 24.69  &        &         &         &        \\ \hline
Spot 2                                          & 21.55  & 15.71   & 37.89   & 24.85  &        &         &         &        \\ \hline
Spot 3                                          & 19.46  & 14.27   & 31.91   & 34.36  &        &         &         &        \\ \hline
Spot 4                                          & 21.66  & 18.39   & 37.37   & 22.58  &        &         &         &        \\ \hline
Spot 5                                          & 20.96  & 24.57   & 37.55   & 16.92  &        &         &         &        \\ \hline
Spot 6                                          & 16.49  & 14.35   & 32.74   & 36.42  &        &         &         &        \\ \hline
Average                                         & 20(2)  & 17(4)   & 36(3)   & 27(7)  & 18(2)  & 9(2)    & 10(3)   & 61(7)  \\ \hline
\end{tabular}
\caption{Quantitative EDS spectra data determining the K, Cu, Te, and O composition at different points of the K$_{2}$CuTeO$_{4}$(OH)$_{2}$ $\cdot$ H$_{2}$O measured sample. The atomic percent of K, Cu, Te, and O is calculated from the reported average weight percent. The calculated ratio of K:Cu:Te is close to 2:1:1:8 in agreement with the compound's stoichiometry.}
\end{table}

\begin{figure*}[h!]
    \centering
    \includegraphics[scale = 0.75]{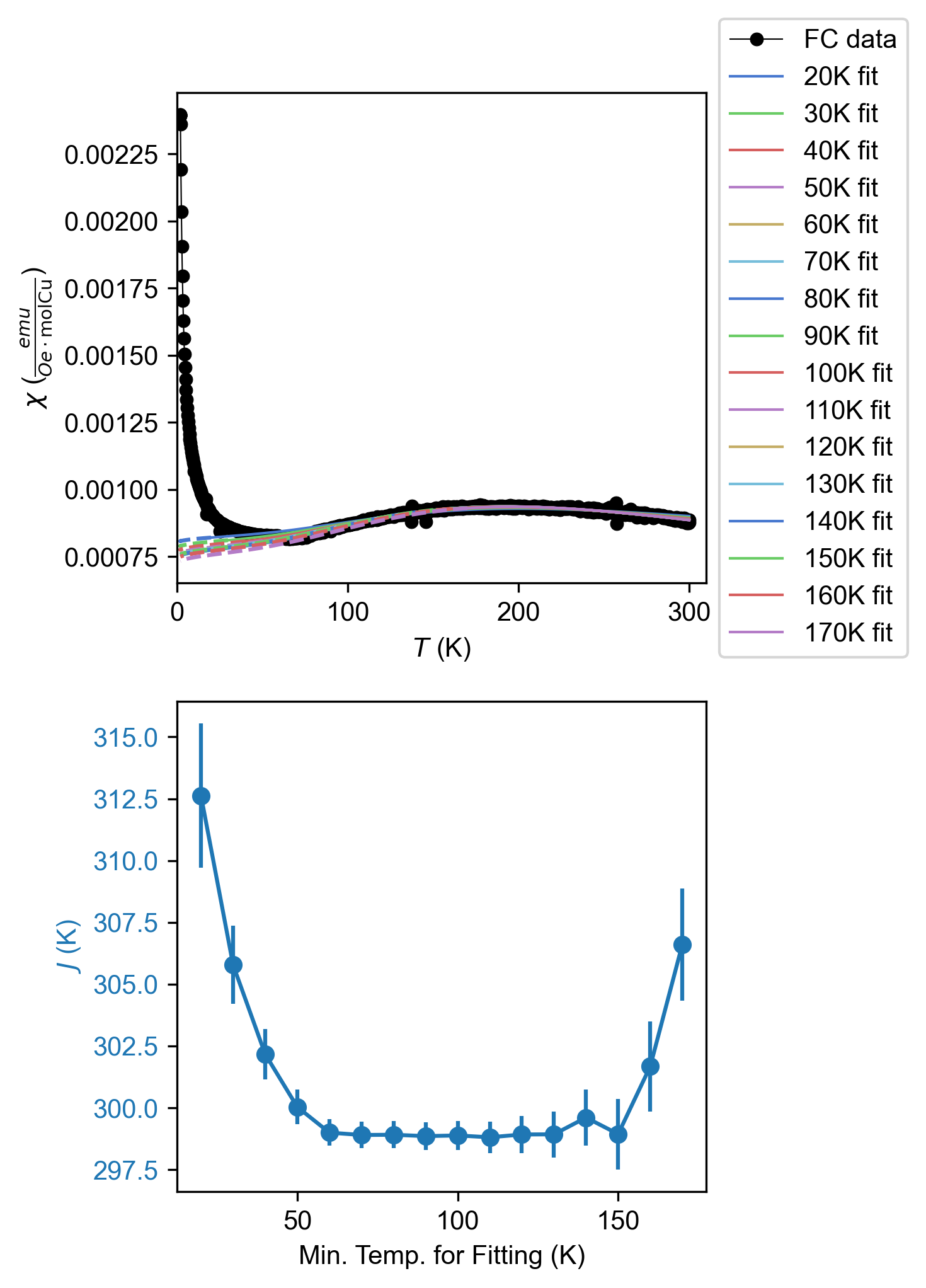}
    \caption{(top) Uniform chain AFM Heisenberg fitting of the magnetic susceptibility of K$_{2}$CuTeO$_{4}$(OH)$_{2}$ $\cdot$ H$_{2}$O choosing different minimum temperature for the high-temperature range of the fitting procedure. (bottom) Resulting fit parameter $J_{max}$ from the choice of different minimum temperature cut-offs in the fitting procedure. Only temperature ranges for which $J_{max}$ remained stable were chosen to be included.} 
\end{figure*}

\begin{figure*}[h!]
    \centering
    \includegraphics[scale = 0.75]{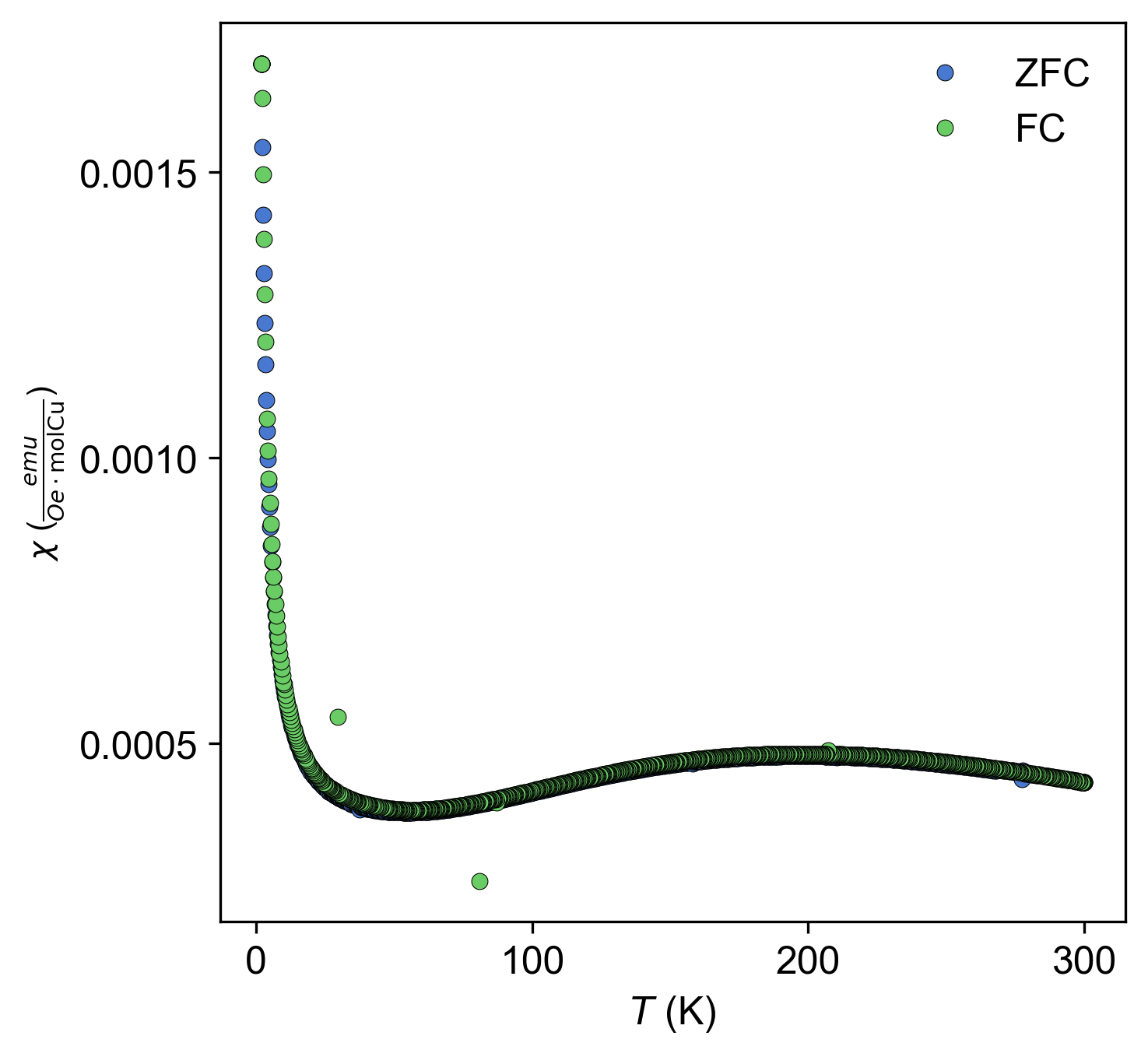}
    \caption{Molar magnetic susceptibility of K$_{2}$CuTeO$_{4}$(OH)$_{2}$ $\cdot$ H$_{2}$O measured in ZFC and FC regimes with an external field of $\mu_{0}H$ = 1 T. } 
\end{figure*}

\begin{figure*}[h!]
    \centering
    \includegraphics[scale = 0.75]{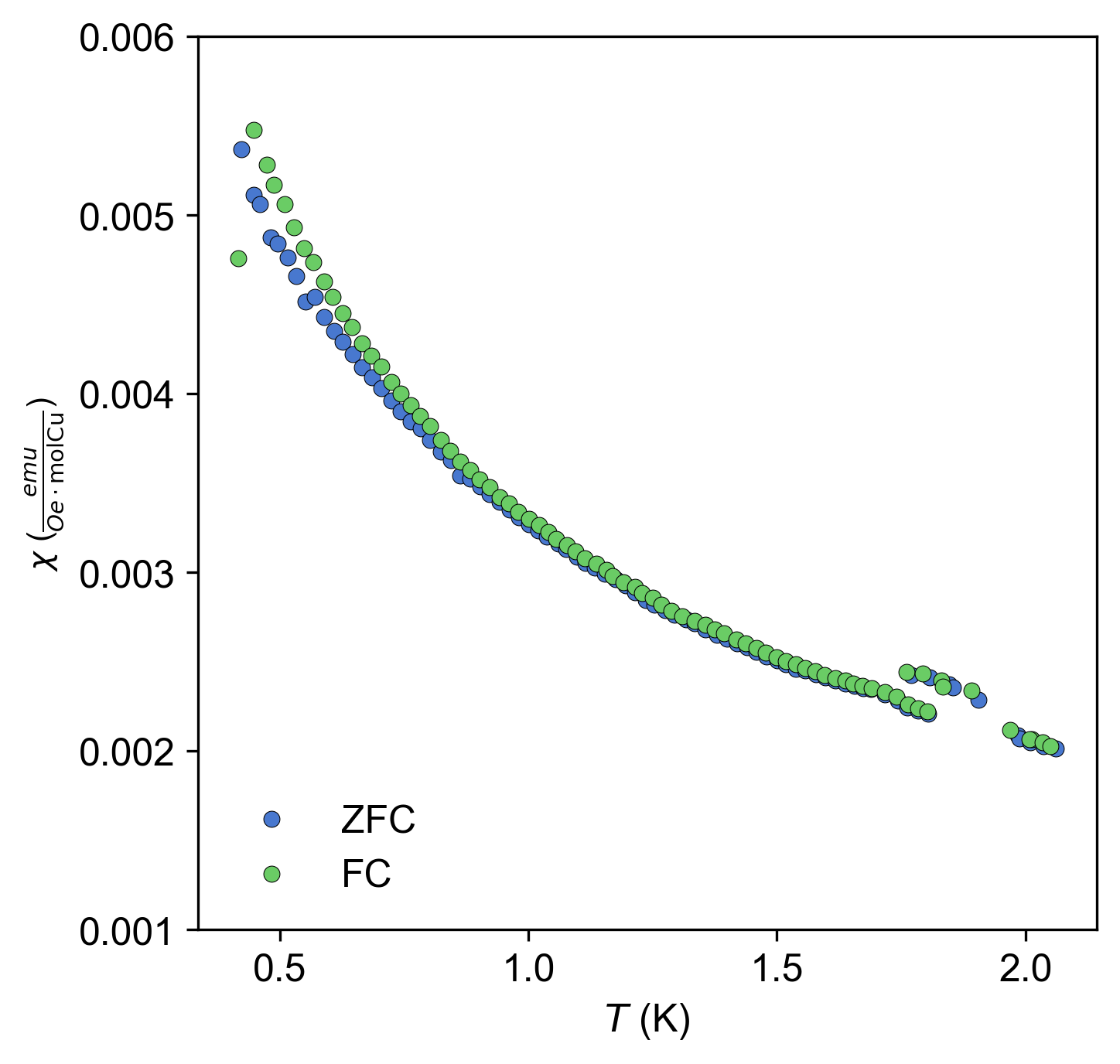}
    \caption{Molar magnetic susceptibility of K$_{2}$CuTeO$_{4}$(OH)$_{2}$ $\cdot$ H$_{2}$O measured in ZFC and FC regimes with an external field of H = 100 Oe down to T = 0.4 K with no magnetic transitions observed.} 
\end{figure*}